\documentclass{article}
\usepackage{geometry,setspace}
\usepackage{amsmath}
\usepackage{amsfonts}
\usepackage{amssymb}
\usepackage{graphicx}
\usepackage{hyperref}
\usepackage{enumerate}
\usepackage[round]{natbib}
\usepackage{marvosym}
\usepackage{wasysym}
\providecommand{\U}[1]{\protect\rule{.1in}{.1in}}

\makeatletter

\newcommand{\distas}[1]{\mathbin{\overset{#1}{\kern\z@\sim}}}

\newtheorem{theorem}{Theorem}

\newtheorem{remark}[theorem]{Remark}

\begin{document}
       
\title{There is Individualized Treatment. \\Why Not Individualized Inference?}
\author{Keli Liu and Xiao-Li Meng\\(\textit{keliliu@stanford.edu \ \& \ meng@stat.harvard.edu})\\Stanford University and Harvard University}
\date{\today}
\maketitle
{\abstract Doctors use~statistics to advance medical knowledge;~ we use a medical
analogy to introduce  statistical inference ``from scratch'' and to highlight an improvement. 
Your doctor, perhaps implicitly, predicts the effectiveness of a treatment  for \textit{you} based on its performance in a clinical trial; 
the trial patients serve as \textit{controls}\textit{} for you\textit{. }The same
logic underpins statistical inference: to identify the best statistical procedure to use for a problem, we simulate
a set of \textit{control problems} and evaluate candidate procedures on the controls.~ 
Now for the improvement: recent interest in personalized/individualized medicine stems
from the recognition that some clinical trial patients are better controls for you than others. 
Therefore, treatment decisions for you should 
depend only on a subset of \textit{relevant} patients. 
\textit{Individualized statistical inference} implements this idea for control problems (rather than patients). Its potential for improving data analysis 
matches personalized medicine's for improving healthcare. 
The central issue---for both individualized medicine and individualized inference---is how to make the right \textit{relevance robustness trade-off:}  
if we exercise too much judgement in determining which controls are relevant,
our inferences will not be robust. How much is too much? We argue that the unknown answer is the Holy Grail  of statistical inference. 

\section*{Prologue: The Data Doctor}

A usual course on statistical inference teaches its \textit{mechanics}: how to
construct confidence intervals or conduct hypothesis tests through the use of
probabilistic calculations. We then become so fluent in the modern language of
statistics that we forget to ask: Why did the language develop this way? Can
inference be done using a different language? For example, why introduce the
notion of probability at all---is it truly indispensable? This article does not
aim to introduce the reader to the language of inference as currently spoken by
many statisticians. Instead, we try to create a language for inference
\textquotedblleft from scratch,\textquotedblright\ and stumble upon the main
statistical dialects (and a few variants) by accident. By from scratch,\ we
mean that we justify each step in the construction employing only
\textquotedblleft common sense\textquotedblright. In fact, we claim that one
only needs to understand the following problem to understand statistical inference.

\bigskip

\textbf{The Doctor's Problem. }A doctor needs to choose a treatment for her
patient, Mr. Payne. Her options are the standard treatment, $A$, and an
experimental treatment, $B$.
What does the doctor do to make her decision? She goes and finds out how $A$
and $B$ worked on \textit{other }patients. Suppose she discovers that
treatment $B$ outperformed $A$ on patients in a large randomized clinical
trial. She wants to apply this result to Mr. Payne. For this to be a
reasonable action, the patients in the clinical trial need to be good
\textit{controls }for Mr. Payne (who is 50 years old, weighs 200 pounds,
exercises once a month, etc.). Of course, she realizes that not all patients in
the trial are good controls (certainly not the 12 year old girl). So she
selects a subset of patients who closely match Mr. Payne's characteristics 
and bases her decision on $A$ and $B$'s performance over this subset.

\begin{figure}[ptbh]
\begin{center}
\includegraphics[width=0.7\textwidth]{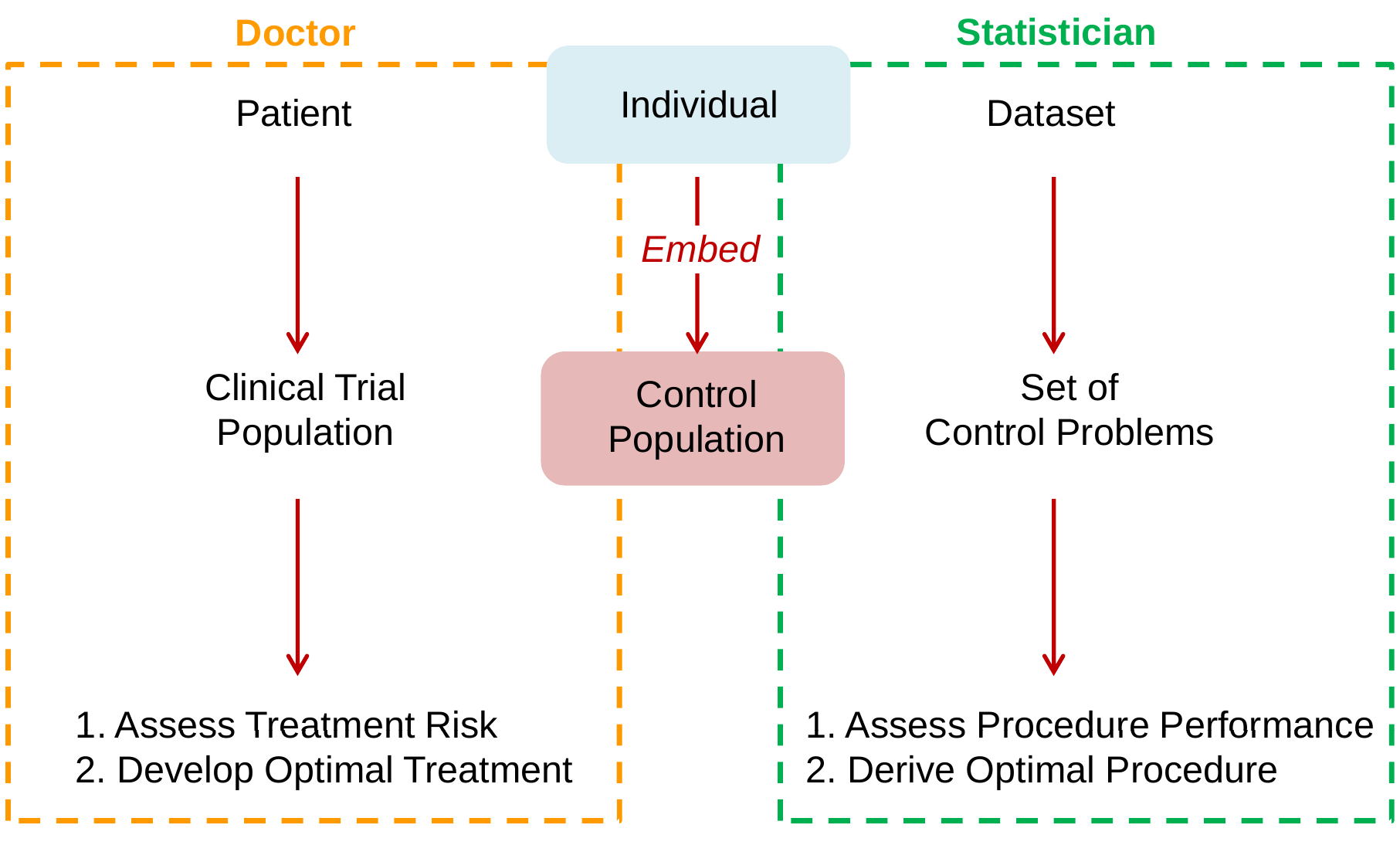}
\end{center}
\caption{The statistician needs to treat his ``patient", a dataset,
with a statistical procedure. He chooses between candidate procedures
based on their performance across a set control problems.}%
\label{fig:doctor}%
\end{figure}

\bigskip

In summary, the doctor does two things: (i) she obtains controls for Mr. Payne
and (ii) among the initial controls, she selects the subset that is most
relevant for Mr. Payne. Statisticians do the same two things; we are the
doctors for data (see Figure \ref{fig:doctor}).  Throughout this review, we exploit 
this analogy to derive insights into
statistical inference. A statistician's patients are datasets for which he
hopes to choose the analysis procedure leading to the most accurate scientific
conclusions. To test the effectiveness of candidate procedures, a statistician
must also set up a clinical trial; he needs to simulate \textit{control
problems} and see how well his procedures perform on the controls. A great
many jobs involve doing something similar: a rocket scientist runs simulations
before launching an actual rocket, a basketball coach runs scrimmages to
simulate a real game. But just as we wouldn't trust a statistician to design
rocket simulations, so there are tricks for getting simulated control problems
to closely match the statistical problem of interest. That is, creating
\textquotedblleft good\textquotedblright\ control problems represents the key
stumbling block in inference.

In our attempt to create good controls, we
will encounter a fundamental tradeoff: \textit{the relevance--robustness tradeoff. }It
is easy to understand in the context of the doctor's problem. If we estimate
Mr. Payne's treatment response by the average response across all clinical
trial patients, we will likely do poorly; patients whose background variables
differ from Mr. Payne's may respond to treatment differently. But if we
predict Mr. Payne's treatment response using only those patients who match Mr.
Payne's background exactly, we will still do poorly. There are too few (or
none) of such patients in the clinical trial; accordingly, our estimate will be too variable. Thus, our
inferences can be hurt either by too much matching (not robust) or too little
matching (not relevant). We will see that this lesson generalizes directly to
the statistical context where we must decide how closely the features of a
control dataset should match those of the actual dataset in order for the
control problem to be deemed relevant.

No consensus exists on the ``right" degree of matching, neither in the doctor's case nor the statistician's. However, it seems intuitively clear 
that no single answer can work for all problems. This maxim is the inspiration for
personalized medicine: for each patient, the doctor begins \textit{anew} the process
of selecting controls, striking a balance
between relevance and robustness that is tailored to the patient's situation. Similarly, we should try to individualize
statistical inferences to the problem at hand. We will argue that this spirit of individualization is
counter to the two standard statistical methodologies---Bayesian and (unconditional) Frequentist inference---which coincide with the
\textit{static} positions of always performing complete matching or never performing
any matching.  The need for personalized medicine is
clear. We hope this article will make a convincing argument that   the need for
\textit{individualized inference }is equally urgent.  

The layout for the remainder of  our article  is as follows:

\begin{description}
\item[Section 1.]  A doctor can go out into the world and find controls for
Mr. Payne. How can a statistician create control problems when he only has one  data set? Anyone who has taken standardized tests knows that
\textquotedblleft problems\textquotedblright\ fall into types. If we can use the observed data to identify the \textquotedblleft type\textquotedblright\ of a statistical
problem, we
can then simulate control problems of the \textit{same type}.

\item[Section 2.] Simulated control problems will not perfectly resemble our actual problem. We should draw inferences
 using only those controls which match the actual problem on important features, e.g., the sample size. 
Through a series of examples, we show
how to weigh gains in our inference's relevance (from matching on a feature) against potential losses in its robustness. 

\item[Section 3.] Many judgments underpin the selection of a set
of ``relevant" controls.  We discuss strategies to make our inferences
robust to \textit{mis}judgments. These
include the use of pivots to create confidence intervals and the minimax
principle. We give a unifying view of these strategies as ways to
\textquotedblleft create symmetry/invariance.\textquotedblright\ 

\item[Section 4.] We map different statistical methodologies according to the
relevance-robustness tradeoff they adopt in creating control problems. We see that Bayesian inference
coincides with complete matching and (unconditional) Frequentist inference with no matching.
Neither extreme seems appealing; we suggest compromises based on
partial matching, which include the controversial  Fiducial inference.
\end{description}

\section{What Makes It a \textit{Statistical} Inference?}

\label{sec:repe}

\subsection{It's the Built-in Uncertainty Assessment ...}

\label{sec:population}

We go to the doctor because, unlike Google, the doctor can distinguish
(hopefully) which treatment among available options is most appropriate
\textit{for me}. Similarly, the statistician must choose the method of
analysis most appropriate for a \textit{particular} dataset. A common logic
governs the doctor's and statistician's choice: to assess the effect of a
procedure (treatment $B$) on a target individual (Mr.\ Payne), we need to
obtain \textquotedblleft controls\textquotedblright\ for that target individual.

\textbf{Control Patients}. Let $\smiley{}$ denote Mr.\ Payne, $y$ his health
outcome under treatment $B$ and $\vec{x}$ his vector of observed background
variables (age, weight, height, etc.). If we could clone Mr. Payne and apply
treatment $B$ to his clone, we would know $y$ perfectly (a
\textit{deterministic} inference). Instead, we settle for a clinical trial of
control patients, for whom we obtain data $\{(\vec{x}^{\prime},y^{\prime})\},$
where $(\vec{x}^{\prime},y^{\prime})$ are the background variables and
response for $\smiley{}^{\prime}$. Throughout, we will use prime notation to
denote aspects of controls and the unprimed versions of the same quantities to
refer to aspects of the target individual. The prime notation reflects our
desire for controls to mimic our target, $\smiley{}$, as closely as possible.

\textbf{Control Problems. }The target \textquotedblleft
individual\textquotedblright\thinspace\ $\smiley{}$, is now a statistical
\textquotedblleft problem\textquotedblright\ which comprises two parts: (i)
the dataset, $D$, being analyzed and (ii) some unknown \textquotedblleft
truth", $\theta$, we hope to infer, which may or may not correspond to a parameter in a statistical model. For concreteness, suppose we observe a
sequence of 9 colored pixels ($D$) and wish to predict the color of the 10th
pixel ($\theta$), as in Figure \ref{fig:iid}b. There are a wealth of
prediction algorithms we could use to accomplish this task. To choose among
them, we evaluate their accuracies on a set of control problems $\left(
D^{\prime},\theta^{\prime}\right)  $. Each control problem is a sequence of 10
pixels. We apply the candidate prediction algorithms to the first 9 pixels ($D^{\prime}$),
and check their outputs against the 10th ($\theta^{\prime}$).

\begin{figure}[ptbh]
\begin{center}
\includegraphics[width=0.7\textwidth]{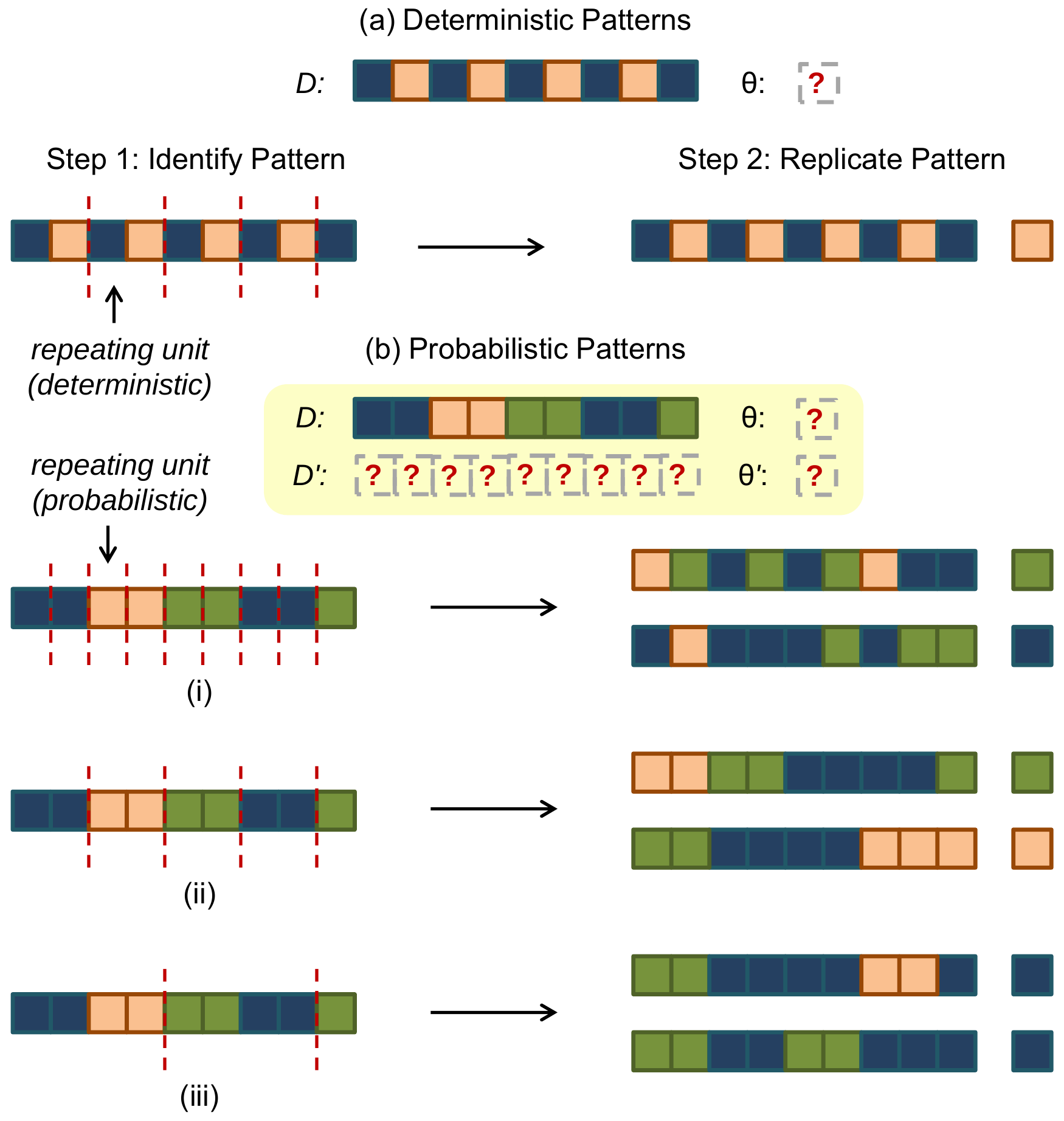}
\end{center}
\caption{Diagram (a) shows a sequence following a \textit{deterministic}
pattern, and diagram (b) shows a sequence constructed from replications of a
\textit{probabilistic} pattern. To create controls, we first decompose the data into the 
repeating subunits of a
pattern. Plots (i), (ii), and
(iii) represent three possible decompositions. We then simulate
controls by randomly sampling (with replacement) from the set of observed
subunits and stringing the sampled subunits together until we obtain 10
pixels or more. To the right of each arrow are two example controls simulated
according to the decomposition on the left.}
\label{fig:iid}
\end{figure}

In our analogy, the statistical problem, $\smiley{}=(D,\theta)$, plays the
role of Mr. Payne, $\smiley{}=\left(  \vec{x},y\right)  $. The doctor observes
$\vec{x}$ and wishes to infer/predict $y$; the statistician observes $D$ and
wishes to infer $\theta$. An important consequence of this parallelism is that
just as we improve the relevance of our inference for Mr.\ Payne by choosing controls which match his background, $\vec{x}$, we should ensure control
problems, $\left(  D^{\prime},\theta^{\prime}\right)  $, for
$\smiley{}=(D,\theta)$ match its \textquotedblleft background", $D$ (see
Section \ref{sec:relevance}). 

\textbf{A Conundrum: Creating Controls from }$D$ \textbf{Alone.} A doctor can
go out into the world and solicit clinical trial subjects. Her controls
therefore contain information \textit{external }to Mr. Payne's observed
characteristics, $\vec{x}$. But typically the \textit{only }input the statistician
has in creating control problems is the data, $D$, supplemented by common sense and 
subject matter knowledge. 
It is absurd to think that one can evaluate the
risks of various treatments based on observing Mr. Payne's background,
$\vec{x}$, alone, i.e. without experimentation on others. Since $\left(
D,\theta\right)  $ parallels $\left(  \vec{x},y\right)  $ in our analogy,
isn't it equally absurd to try and conduct inference using only $D$? Up to
this point, our reasoning process has closely paralleled the doctor's. It is
here that statistics makes a wholly original contribution in supplying a logic
for creating meaningful controls $(D^{\prime},\theta^{\prime})$ from $D$
alone. Not only do we possess the ability to predict $\theta$
from $D$, but we can also assess the accuracy of our prediction using only $D$.
The capacity to assess uncertainty is \textit{built into} the data. This is the \textit{magic }of statistics. 
Of course, all magics come with ``hidden designs," which we will reveal below.  

\subsection{The Data As A Replicating Pattern}

\label{sec:repeat}
The pixel example (Figure \ref{fig:iid}) hints at a strategy for
creating controls $\left(  D^{\prime},\theta^{\prime}\right)  $ based only on
$D$. To make it explicit, consider a simpler problem (depicted in Figure \ref{fig:iid}a) 
where we \textit{assume} the sequence follows a deterministic
pattern. From the observed data, we easily identify the repeating unit of this pattern as a block of blue-orange pixels. 
Using this fact, we can simulate controls which follow the same pattern
as the target. At a high level, our strategy for creating controls can be summarized as follows:

\begin{enumerate}
\item Assume the target problem, $\left(  D,\theta\right)  $, follows a pattern.

\item Extract/estimate the pattern using only the observed data, $D$.

\item Produce controls, $\left(  D^{\prime},\theta^{\prime}\right)  $, based
on the extracted pattern.
\end{enumerate}
For the deterministic pattern in Figure \ref{fig:iid}%
a, this strategy would lead to perfect controls: $\left(  D^{\prime}%
,\theta^{\prime}\right)  =\left(  D,\theta\right)  $.

\textbf{Probabilistic Patterns. }A \textit{deterministic} pattern would not be
useful in describing the (seemingly) more complex sequence in Figure
\ref{fig:iid}b. So we extend our strategy: use a \textit{probabilistic
}pattern to describe the color scheme of the first 9 pixels and assume that
the 10th pixel follows the same pattern. A probabilistic pattern is typically captured by a (probability) \textit{distribution}---it is a pattern for \textit{how things
vary}. For example, when we say that height in a human population follows a
normal distribution, we mean that \textit{variation} in height follows the
pattern of a bell shaped curve. In the current problem,  we can try to describe the \textit{variation }in pixel
colors using a distribution, call it $f$. For example, based
on the 9 observed pixels (4 blue, 3 green, 2 orange), we might guess that each
individual pixel has probability 4/9, 3/9 and 2/9 respectively of being blue,
green or orange (independent of the other pixels). This constitutes an
estimate $\hat{f}$ of $f$. We can then produce control sequences, $\left(  D^{\prime
},\theta^{\prime}\right)  $, according to the estimated pattern $\hat{f}$; the
two example controls on the right of Figure \ref{fig:iid}b(i) were simulated
in this way.

\textbf{Robustness versus Descriptive Power}. The above pattern assumes that
the colors of individual pixels are independent of each other. This pattern fails
to capture the block structure of the sequence: the first two colors are the
same, so are the next two, and so on. To fix this
deficiency, we can choose \textit{blocks} of pixels as our repeating unit, as
in Figure \ref{fig:iid}b(ii) and (iii). That is, color variation, not for a
single pixel but for a block of pixels, follows some pattern $f$. Our
estimated pattern $\hat{f}$ will then encode features of the observed sequence that
are specific to pixels \textit{as well as} those specific to blocks; see the
 controls corresponding to decompositions (ii) and (iii) in Figure
\ref{fig:iid}b.

With this greater descriptive power comes the danger that our pattern captures
features of the observed pixels that do not generalize (artifacts).  Under decomposition (iii),  we observe two blocks of size 4: a block of
blue-blue-orange-orange (bboo) pixels and a block of green-green-blue-blue
(ggbb) pixels. We might guess that each non-overlapping block of 4 pixels is
bboo or ggbb with probability $1/2$. This choice of $\hat{f}$ produces
controls with a rather curious feature: a two pixel orange block \textit{must
be} preceded by a two pixel blue block. Prediction algorithms which exploit
this feature would do well on our simulated controls; they would do poorly on
our actual problem if such a feature turns out to be an artifact.  The point is,
we have no way to assess whether such a feature is genuine or not since a
block of orange pixels appears only once in the observed data. If we had
further replications of 4 pixel-blocks in our observed data, we could
specify $\hat{f}$ using only a portion of the replications, and then
assess the generalizability of $\hat{f}$ by seeing how well $\hat{f}$ describes the remaining replications.
The more replications we have for testing $\hat{f}$, the better we can ensure
that whatever $\hat{f}$ we use encodes minimal artifacts.

This sets up a tradeoff between descriptive power and robustness.
Decomposition (i) lacks the descriptive power to capture across pixel
dependencies in pixel color; decomposition (iii) captures features of the
observed sequence which may simply be noise. Creating reliable controls hinges
on identifying the appropriate decomposition of the data into replicating
subunits \citep[see][]{kruskal1988}. Statistical inference generalizes
reasoning about deterministic patterns to reasoning about
\textit{probabilistic }patterns. The fundamental requirement of a
pattern---something that repeats---remains unchanged.

\textbf{No Free Lunch. }To summarize, by decomposing $(D,\theta)$ into basic
building blocks which follow a probabilistic pattern we can create controls
$\left(  D^{\prime},\theta^{\prime}\right)  $ from $D$ alone. This strategy
turns on an \textit{intestable} assumption: the pattern underlying $D$
persists through $\theta$. That is, the pattern present in the 9
\textit{observed }pixels persists through the 10th and \textit{unobserved
}pixel. This assumption is the price we pay for creating
\textit{internal }controls (those based on $D$ alone). In contrast, whenever the doctor uses
clinical data, she uses \textit{external controls }for Mr. Payne. To see the
difference clearly, imagine you want to study yourself. You can experiment on
other humans like you (an external source), or you can take some cells from
your body (an internal source), experiment on them and extrapolate the
conclusions using biological knowledge of how you are (or are not) the sum of
your cells---the latter clearly requires more assumptions.
        
\begin{figure}[ptbh]
\begin{center}
\includegraphics[width=0.6\textwidth]{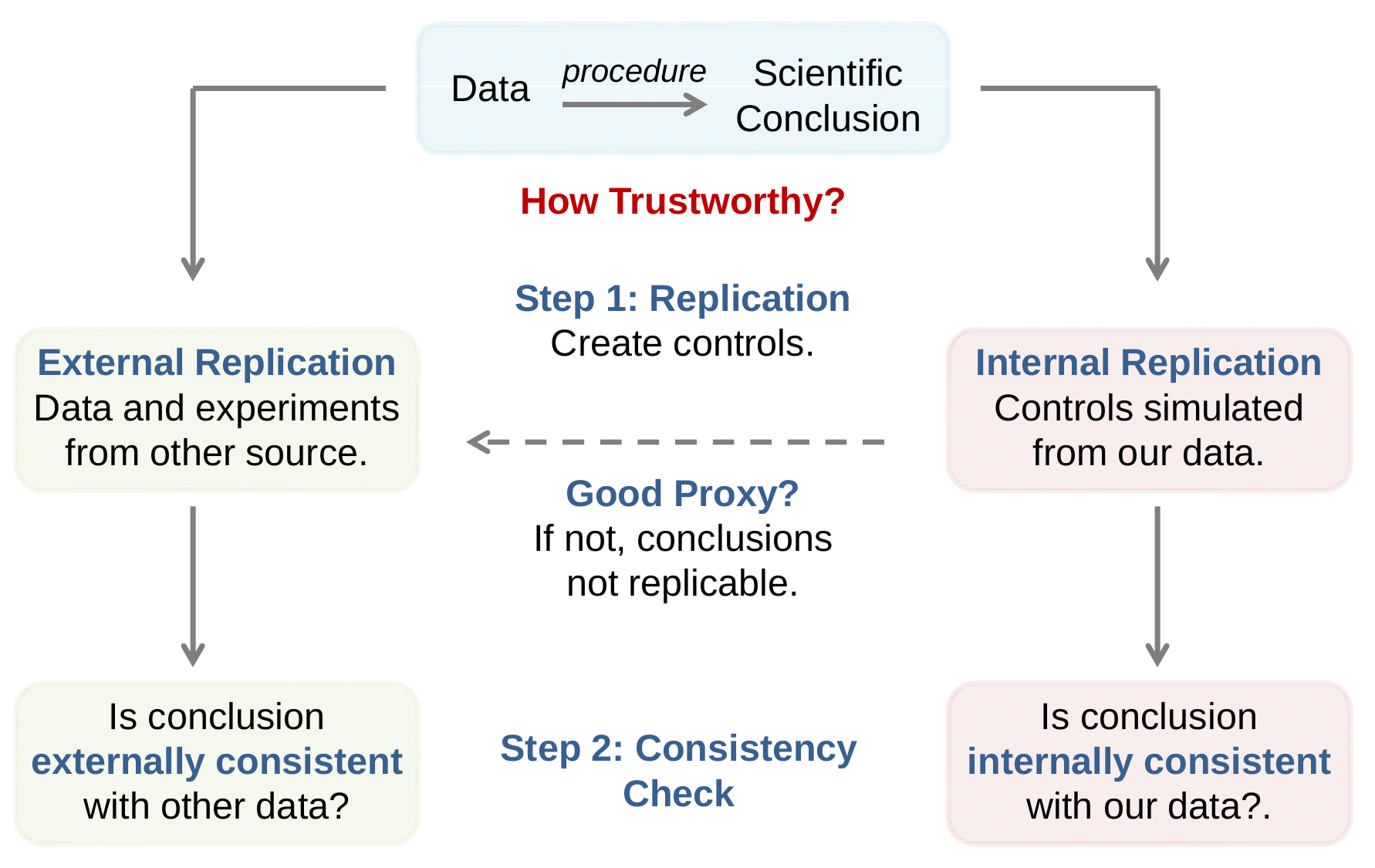}
\end{center}
\caption{The diagram depicts how one can go about assessing the validity of a
scientific conclusion drawn from the data. The gold standard for validation is
to replicate the experiment, producing new data, and see whether conclusions
drawn from the replicate data are consistent with the current conclusion.
Statistical models give us a way to simulate replicate data without having to
perform new experiments. If these simulations produce data which closely match
those generated by genuine experimental replications, then statistical
significance says something meaningful about the real life reproducibility of
scientific findings. Otherwise, a statistical uncertainty assessment amounts
to only an internal consistency check, i.e., it answers whether the conclusion
is reasonable given the current data, not whether the conclusion generalizes.}%
\label{fig:replication}%
\end{figure}

External replications are a cornerstone of the scientific method. To test the
validity of a conclusion, an experiment is repeated over and over to see if
consistent conclusions can be obtained. The \textit{Big Idea} from statistics
is the construction of internal controls, giving the data a \textit{built-in
capacity} to assess the uncertainty of scientific conclusions drawn from it.
That is, having performed this experiment, and before I perform validation
experiments, how confident should I be about my conclusions? As discussed
above, internal controls are usually of lower quality than external controls
because they entail additional assumptions. When
uncertainty assessments based on internal replications diverge from those
based on external replications, a failure in statistical modelling has
occurred, leading to unreplicable research (see Figure \ref{fig:replication}).
Hence the million dollar question is how to create internal replications which
closely resemble external replications.

\subsection{Frequent Misconceptions: The Meaning of A Statistical Model}

\label{sec:misconception} This section aims to clarify what we assume \textit{and do not
assume} when simulating controls according a probabilistic pattern/distribution.
In the literature, this modeling assumption usually takes the form:
\textquotedblleft The data \textit{come from} such and such
distribution.\textquotedblright\ This phrasing may give the false impression that...

\begin{description}
\item[Misconception 1.] A probability model must describe the
\textit{generation} of the data.
\end{description}

A more apt description of the model's job (in inference) is \textquotedblleft
Such and such probabilistic pattern produces data which \textit{resemble} ours
in important ways.\textquotedblright\ To create replicas (i.e., controls) of
the Mona Lisa, one does not need to bring da Vinci back to life ---a camera
and printer will suffice for most purposes. Of course, knowledge of da Vinci's
painting style will improve the quality of our replicas, just as scientific
knowledge of the true data generating process helps us design more meaningful
controls. But \textit{for purposes of uncertainty quantification}, our model's
job is  to specify a set of controls that \textit{resemble} $(D,\theta)$.
Nowhere is this point clearer than in applications involving computer
experiments where a probabilistic pattern is used to describe data following a
known (but highly complicated) deterministic pattern
\citep[][]{kennedy2001,conti2009}. We need a \textit{descriptive} model,
not necessarily a \textit{generative} model. See \citet{lehmann1990},
\citet{breiman2001} and \citet{hansen2001} for more on this point. 

\begin{description}
\item[Misconception 2.] Because we use a probabilistic model to simulate
controls $(D^{\prime},\theta^{\prime})$, we must have assumed that $D$ and
$\theta$ are \textit{random}.
\end{description}

Probability and randomness, so tightly yoked in our minds, are in fact
distinct concepts. The language of probability supplies a convenient way to
\textit{represent} a set of controls. If we decide that $(D_{1},\theta
_{1}),...,(D_{10},\theta_{10})$ act as ten equally compelling controls for
$(D,\theta)$, we can mathematically represent this set of controls as a
uniform distribution over these ten problems. Nothing has been said about $D$
or $\theta$ being random! Randomization inference (see Section
\ref{sec:diagonal}) is a rare instance where the randomness described in the
probability model actually corresponds to a physical act of randomization. We
can certainly motivate our choice of control problems by conducting
thought experiments such as \textquotedblleft What if the pixels were
generated by a random mechanism?\textquotedblright. But at the end of the day,
probability is essentially a tool for bookkeeping, just like the abacus.

\subsection{Modeling Complex Datasets by Layering Probabilistic Patterns}

\label{sec:similarity}

In Section \ref{sec:repeat} we decomposed our data into independent
replications of a single probabilistic pattern---an i.i.d., \textit{independent and identically distributed},  model. How do we
generalize this idea to capture more complicated data structures? An artist
often creates a painting through a series of layers---the backmost layer
captures broad shapes and general atmosphere (global features) whereas the
foremost layers capture fine detail (local features). We can use a similar
layering strategy: for us, each \textquotedblleft layer\textquotedblright%
\ will be a probabilistic pattern, tailored towards capturing data variation at a
certain \textit{resolution }level. This is a fundamental idea in many applications, e.g., wavelets \citep[see][]{daubechies2010,donoho1995}.

\begin{figure}[ptbh]
\begin{center}
\includegraphics[width=0.7\textwidth]{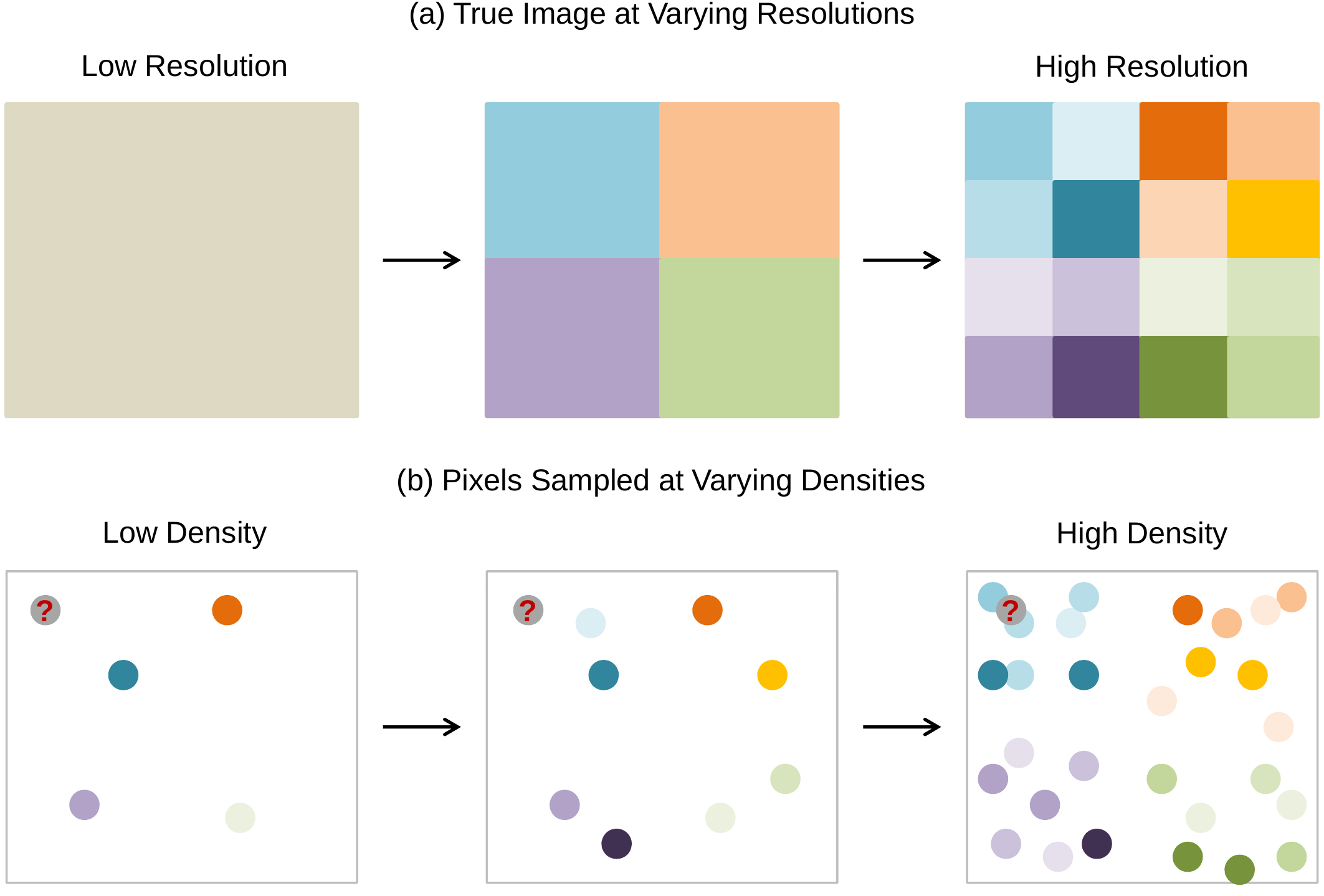}
\end{center}
\caption{(a) shows the color of the true image averaged over the entire square (left panel), averaged over subsquares (middle panel), or averaged over sub-subsquares (right panel). The true image exhibits smoothness in color variation. (b) shows three datasets (of varying
sampling density) comprised of pixels. Each pixel is assumed to be ``area-less''.}%
\label{fig:resolution}%
\end{figure}

To see clearly how the artist analogy applies, suppose we
have an image, thought of as an infinite set of pixels, each of which is
\textquotedblleft area-less\textquotedblright. Our data comprises of the color
$y_{i}$ (a length 3 vector specifying the RGB values) and position $p_{i}$
(the cartesian coordinates) of $n$ randomly sampled pixels; the goal is to
impute the color, $y$, of a pixel at a target location $p$ (see Figure
\ref{fig:resolution}b). So $D=\left\{  \left\{  p_{i},y_{i}\right\}
_{i=1}^{n},p\right\}  $ and $\theta=y$.

\begin{itemize}
\item \textit{A Resolution 1 Model. }Following Section \ref{sec:repeat}, we
can model color variation at the pixel level using an i.i.d. model, i.e., we
assume that the color of each pixel follows some probability distribution
\textit{independently} of all other pixels. This model completely ignores the spatial properties of our image:
pixels that are closer together are more likely to share a similar color.
Control images simulated by this model will exhibit no spatial smoothness.
Hence, algorithms which exploit such smoothness properties will perform poorly
on our controls even if they perform well on our actual image.
\end{itemize}

The layering strategy can help us encode smoothness properties into our
control images. The model we use below is admittedly naive and has some fatal
flaws but its simplicity gets our key points across.

\begin{itemize}
\item \textit{A Resolution 2 Model. }Let us divide our image into 4 subsquares
as in the mid-panel of Figure \ref{fig:resolution}a. We decompose color
variation (across pixels) into two sources: variation in color across the 4
subsquares and variation in color within each subsquare. We assume that the first
type of variation follows a probabilistic pattern, $f^{\left(  1\right)  }$,
and the second type of variation follows a probabilistic pattern $f^{\left(
2\right)  }$---this constitutes a resolution 2 model. If in our actual image all pixels in the same subsquare share the
same color,  then  the resolution 2 model would easily pick up on this feature, but
the resolution 1 model would not. 
\end{itemize}

Despite being an improvement,  the resolution 2 model  commits the same
sin as the resolution 1 model but just on a finer scale: it will not be able to capture spatial properties
of our image \textit{within} each subsquare.  Of course, we can further increase our model resolution by subdividing each subsquare into 4
\textquotedblleft sub-subsquares\textquotedblright\ (right panel of Figure
\ref{fig:resolution}a) and employ a resolution 3 model. Continuing in this manner, we
can build models of arbitrarily high resolution $R$.
Since higher resolution models offer us greater descriptive power, one might
be tempted into choosing $R=\infty$. However, we run into the same problem
encountered in Section \ref{sec:repeat}: descriptive power comes at the cost
of robustness. Highly flexible models tend to \textquotedblleft
overfit\textquotedblright\ the data, i.e., our simulated controls will exhibit
idiosyncratic   features\ specific to the
 pixels we happen to observe. Our choice of model resolution, $R$, should be compatible with the ``data resolution." For example, to meaningfully estimate how smooth our image is,
we must sample pixels at sufficiently high density--- Figure
\ref{fig:resolution}b shows three separate datasets with different sampling
densities. If we sampled our image at low density (left panel of Figure
\ref{fig:resolution}b), we have no information to  assess how smooth the image is
within each subsquare. 

In summary, \textquotedblleft layering\textquotedblright\ is a widely
applicable strategy for modeling complex datasets. A prominent
example of its use in statistics is in developing hierarchical models
\citep{gelmanhill,gelman2013}. By combining many layers of probabilistic
patterns, we maintain modeling flexibility while retaining the simplest of
building blocks---the i.i.d. model---within each layer. As in the basic
setting of Section \ref{sec:repeat}, a key challenge is balancing
a model's descriptive capacity and its robustness. 

\bigskip

\begin{remark}
Here, our data are pixels of a \textit{fixed }image. There is nothing random about the
generation of the data. Yet nonetheless, we can describe the data in terms of
\textit{probabilistic }patterns for the purpose of creating control images
$(D^{\prime},\theta^{\prime})$. This reinforces our point in Section
\ref{sec:misconception} that the probabilistic nature of statistical models
has little to do with \textquotedblleft randomness\textquotedblright\ in the
data generating process.
\end{remark}

\subsection{The Weakest Links}

\label{sec:weaklink}

This section identifies  which components of our model are hardest to
specify based on the data alone. In Section \ref{sec:robustness}, we will 
develop inferential strategies which are robust to these weak links. For
concreteness, we first translate the resolution 3 model discussed above into
a mathematical form. Let $\mu^{\left(  0\right)  }$ be the \textquotedblleft
base\textquotedblright\ color of our actual image (Figure \ref{fig:resolution}%
a left panel); the superscript indicates that this is layer 0 in our
multi-layer model. Let $\left\{  \mu_{j}^{\left(  1\right)  }\right\}
_{j=1}^{4}$ be the average color over each of the 4 subsquares, $\left\{
\mu_{k}^{\left(  2\right)  }\right\}  _{k=1}^{16}$ be the average color over
each of the 16 sub-subsquares, and $y_{i}$ be the color of a sampled pixel. We
decompose pixel color into the following 3 sources of variation:%
\[
y_{i}=\mu^{(0)}+\qquad\underbrace{(\mu_{j(i)}^{(1)}-\mu^{(0)})}%
_{\substack{\text{variation across}\\\text{subsquares}}}+\qquad
\underbrace{(\mu_{k\left(  i\right)  }^{\left(  2\right)  }-\mu_{j(i)}^{(1)}%
)}_{\substack{\text{variation across}\\\text{sub-subsquares}}}+\qquad
\underbrace{\left(  y_{i}-\mu_{k\left(  i\right)  }^{\left(  2\right)
}\right)  }_{\substack{\text{variation within}\\\text{sub-subsquares}}}\text{.}%
\]
Here $j\left(  i\right)  $ denotes the index of the subsquare and $k\left(
i\right)  $ the index of the sub-subsquare containing pixel $i$. The
resolution 3 model assumes that the base color $\mu^{\left(  0\right)  }$
follows a distribution $f^{\left(  0\right)  }$, the variations
$\mu_{j}^{(1)}-\mu^{(0)}$ (for $j=1,...,4$) follow a distribution $f^{\left(
1\right)  }$, $\mu_{k\left(  i\right)  }^{\left(  2\right)  }-\mu_{j\left(
i\right)  }^{\left(  1\right)  }$ a distribution $f^{\left(  2\right)  }$, and
$y_{i}-\mu_{k\left(  i\right)  }^{\left(  2\right)  }$ a distribution
$f^{\left(  3\right)  }$. How much information is there in the data for
specifying the probabilistic patterns $f^{\left(  0\right)  },f^{\left(
1\right)  },f^{\left(  2\right)  }$ and $f^{\left(  3\right)  }$ respectively?
Below, we assume that pixels are sampled randomly so that our sample is
\textquotedblleft representative\textquotedblright\ of the overall image.

\begin{itemize}
\item \textit{The highest resolution pattern (\textquotedblleft noise
distribution"): }$f^{(3)}$. Suppose we estimate the average color of a
sub-subsquare, $\mu_{k}^{(2)}$, using the average color of
the sampled pixels in that sub-subsquare, $\bar{y}_{k}^{\left(  2\right)}.$  The \textquotedblleft
residuals\textquotedblright\ $y_{i}-\bar{y}_{k}^{\left(  2\right)  }$ should
approximately follow the distribution $f^{\left(  3\right)  }$ (assuming we
have sampled enough pixels from each sub-subsquare); if in total we sample $n$
pixels from our image, we essentially observe $n$ repeats of the probabilistic
pattern $f^{\left(  3\right)  }$. As a result, it is quite easy to detect
misspecification of $f^{\left(  3\right)  }$ using the data. For example,
residual plots are often used to diagnose whether a normal distribution is
appropriate for describing various aspects of the data.

\item \textit{The lowest resolution pattern (``prior
distribution"): }$f^{(0)}$. Suppose we estimate $\mu^{\left(  0\right)  }$
using $\bar{y}^{\left(  0\right)  }$, the average color across all sampled
pixels, which is an approximate replication of the
pattern $f^{\left(  0\right)  }$. Just as we used the residuals $y_{i}-\bar
{y}_{k}^{\left(  2\right)  }$ to check whether a  specification of
$f^{\left(  3\right)  }$ is reasonable, so we can use $\bar{y}^{\left(
0\right)  }$ to assess the fit of  $f^{\left(  0\right)  }$
\citep[see][]{evans2006}. The key difference of course is that $f^{\left(  0\right)  }$ repeats only once, and hence it cannot be reliably specified  based on the data alone.
\end{itemize}

For medium resolution patterns, there exist no consensus guidelines regarding
the number of replications needed for a robust specification. For example,
there are 16 across sub-subsquare differences (corresponding to $\mu
_{k}^{\left(  2\right)  }-\mu_{j}^{\left(  1\right)  }$) which are
approximate replications of $f^{\left(  2\right)  }$. Is this enough to
meaningfully specify $f^{\left(  2\right)  }$? See \citet[p.62]{cox2006} and
\citet[p.119]{gelman2013} for some discussion. The key takeaway is that in a
multi-layer model, the \textquotedblleft higher resolution\textquotedblright%
\ patterns---which repeat more often throughout the data---are easier to
specify than the \textquotedblleft lower resolution\textquotedblright\ patterns.

A natural question then arises: Can we make our inference
robust to (ideally, independent of) those patterns which are hard to specify?
For example, we might seek a prediction algorithm which performs well on
controls $(D^{\prime},\theta^{\prime})$ \textit{regardless} of the low
resolution pattern $\hat{f}^{(0)}$ used in simulating those controls. This is
a primary goal of Frequentist statistics \citep[see][]{berger1985b,efron1986}.
As we shall see, such robustness\ often comes with a price---loss of
\textit{relevance}. The key challenge in statistical inference---as in the
doctor's problem---is then to balance robustness and relevance when creating
controls. This topic consumes the remainder of this article; we will study it
through an idealized problem detailed below.

\bigskip

\textbf{Problem Setup: }For simplicity, consider data
sampled independently from $n$ units, $D=\left\{  y_{i}\right\}  _{i=1}^{n}$.
We will assume that a simple decomposition of the data 
suffices in describing its structure: $y_{i}=g\left(
\theta,\varepsilon_{i}\right)  $, where $g$ is a known function, $\theta$
varies according to a low resolution pattern $f_{\theta}$ and $\varepsilon
_{i}$ are independent replications of a high resolution pattern
$f_{\varepsilon}$. For
example, $\left\{  y_{i}\right\}  _{i=1}^{n}$ may be multiple measurements for
a quantity of interest $\theta$ with additive error $\varepsilon_{i}$:
$y_{i}=\theta+\varepsilon_{i}$. Note that we use $\theta$ here as both a
parameter in our description of the data and as our quantity of interest
because these two often coincide in practice. To perform inference, we need to
simulate control problems $\left(  D^{\prime},\theta^{\prime}\right)  $ using
specifications $\hat{f}_{\theta}$ and $\hat{f}_{\varepsilon}$. The wrinkle in
our plan is that there is often insufficient information to
reliably specify $f_{\theta}$. Despite the simplicity of this setup, it
allows us to access nearly all the important lessons. 

\bigskip

\begin{remark}
Meaningful specification of $f_{\theta}$ usually requires information external
to the data. As a result, $f_{\theta}$ is often called a \textquotedblleft
prior distribution\textquotedblright\ (it encodes prior/external knowledge).
Despite the special nomenclature,  the
functional purpose of $f_{\theta}$ and $f_{\varepsilon}$ is the same: to
simulate controls. What separates them is the extent to which the data contain
information for specifying $f_{\theta}$ versus $f_{\varepsilon}$. This distinction
is a matter of degree. It is counterproductive to treat $f_{\theta}$ and
$f_{\varepsilon}$ as categorically different objects since there are an increasing
number of datasets which do contain meaningful information for specifying
$f_{\theta}$; efficient methods for such data require treating $f_{\theta}$ on
a more equal footing with $f_{\varepsilon}$ (see Empirical Bayes in Section
\ref{sec:diagonal})
\end{remark}

\section{The Relevance-Robustness Tradeoff}

\label{sec:tradeoff}

\subsection{How Good Are Our Controls?}

\label{sec:questions}

 Let $\hat{\Omega}=\left\{
\left(  D^{\prime},\theta^{\prime}\right)  \right\}  $ denote our set of
simulated control problems, as described in the previous section. The controls in $\hat{\Omega}$ allow us to evaluate the
effectiveness of various statistical procedures in the same way that medical
procedures are tested through clinical trials. The evaluation process has two
basic steps:

\begin{enumerate}
\item \textit{Define the evaluation criterion}. We will use $\Delta^{\prime}$
to denote the loss/error of a procedure when applied to a control problem
$\left(  D^{\prime},\theta^{\prime}\right)  $. For example, we may want to
test whether a hypothesis $H_{0}$ about $\theta^{\prime}$ is true ($T^{\prime
}=1)$ or false $(T^{\prime}=0)$.   A statistical test then takes data as input and
outputs an estimate $\hat{T}^{\prime}$ of the truth value $T'$. Our test errs
($\Delta^{\prime}=1$) whenever $T^{\prime}\neq\hat{T}^{\prime}$; otherwise
$\Delta^{\prime}=0$.

\item \textit{Estimate the procedure's error, }$\Delta$\textit{, for the
target problem, }$\left(  D,\theta\right)  $\textit{, using a summary of its
performance over control problems.} For example, we might estimate $\Delta$ by
the procedure's \textit{average error} over  $\hat{\Omega}$, denoted $\bar{\Delta}^{\prime}$. In hypothesis testing, $\bar{\Delta}^{\prime}$
is  the proportion of controls where the test fails, $\hat{T}^{\prime
}\neq T^{\prime}$.
\end{enumerate}
Figure \ref{fig:errortable} summarizes the notation and terminology associated
with this evaluation process for the three most common inferential tasks:
point estimation, set estimation, and hypothesis testing.

\begin{figure}[ptbh]
\begin{center}
\includegraphics[width=0.9\textwidth]{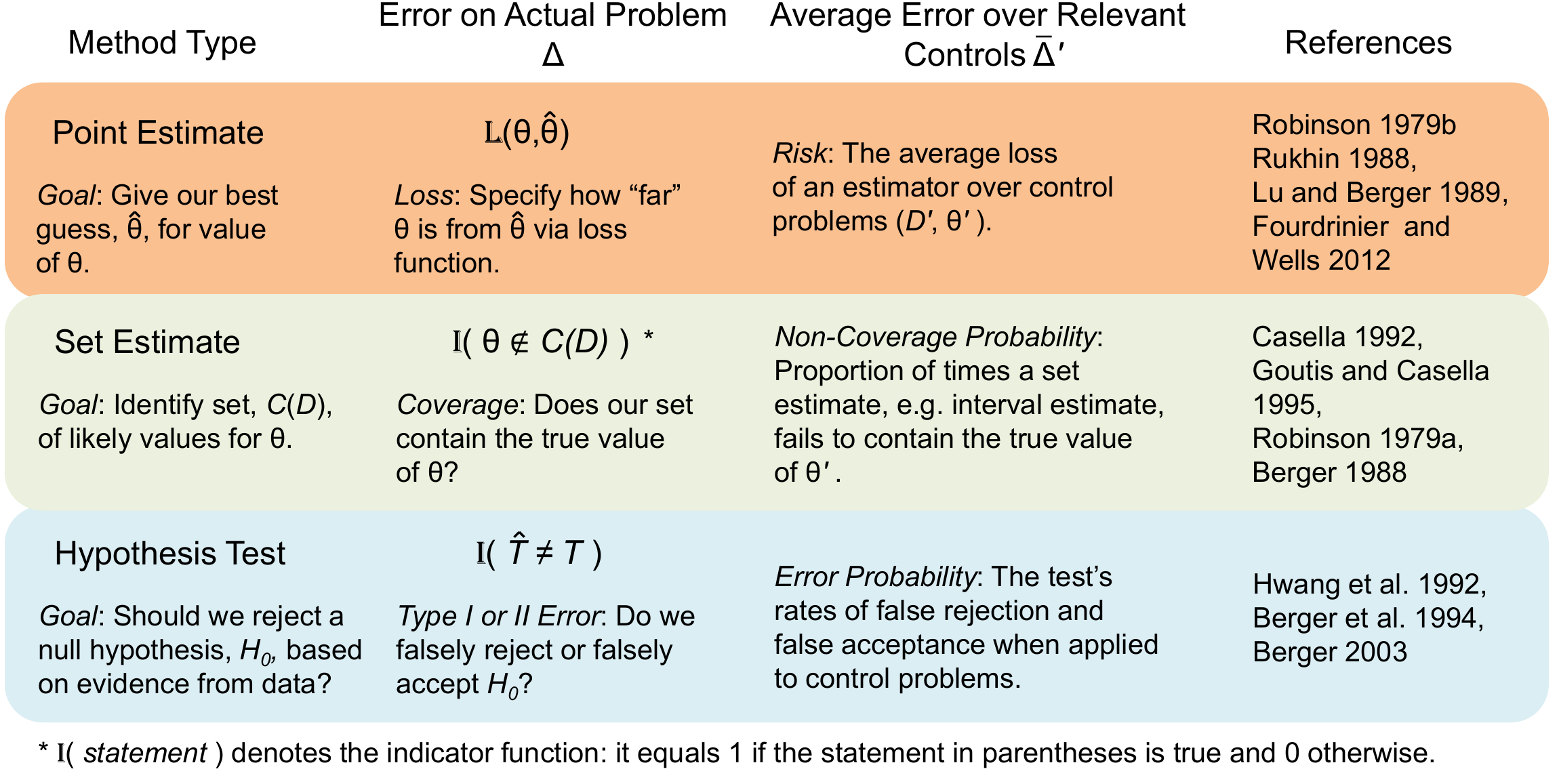}
\end{center}
\caption{Three primary ways to summarize information about $\theta$ are point
estimates, set estimates, and hypothesis tests. For a summary to be
meaningful, we need to assess its quality when applied to the target problem
$(D,\theta)$, i.e., we need to estimate $\Delta$. We do so by evaluating the
performance of each procedure over control problems. The references listed are a
small sample of articles discussing how the relevance of our control problems
affects the accuracy of the resulting estimates $\bar{\Delta}^{\prime}$ of
$\Delta$.}%
\label{fig:errortable}%
\end{figure}
A natural question arises at this juncture: How well does a procedure's
average performance over controls predict its actual performance $\Delta$
\citep[see][]{kiefer1976,kiefer1977,brown1978,sundberg2003}? For example,
suppose a researcher finds a formula for a \textquotedblleft95\% confidence
interval\textquotedblright\ and uses it to produce an interval estimate for a
quantity $\theta$. Is 95\% a good estimate for whether his interval actually
contains $\theta$? The answer obviously depends on the control problems used to compute the  \textquotedblleft%
95\%\textquotedblright\ figure. 

As foreshadowed by our medical analogy, the initial set of controls,
$\hat{\Omega},$ usually contains problems that are a poor imitation of
$\left(  D,\theta\right )$. We can usually select a subset
$\hat{\Omega}_{\text{rele}}\subset\hat{\Omega}$ of controls that more closely
\textit{match} the observed features of $\left(  D,\theta\right)  $, hence is
more \textit{relevant}. This idea has a long history: \citet{fisher1959}
described it as finding a ``reference set" for our problem and proposed the criterion of
``recognizable subsets" \citep[see][Section 7.2]{zabell1992} to judge whether more matching is needed. While conceptually simple, we quickly run into trouble when hammering out the details of the matching process. In particular, the relevance of a control problem turns out to be highly sensitive to the specification of $\hat{f}_{\theta}$ in our model (Section \ref{sec:relevance})---which we
know from Section \ref{sec:weaklink} can be quite unreliable. Since gains in
relevance may be washed out by loss of robustness, we need to strike the right
balance. For example, the reader may be aware that there are 95\%
\textquotedblleft Frequentist\textquotedblright\ confidence intervals and 95\%
\textquotedblleft Bayesian\textquotedblright\ confidence intervals---both have
a success rate of 95\% but with respect to \textit{different }choices of
control problems. In particular, the Bayesian favors relevance and matches
(much) more than the Frequentist. Which of the two made the better tradeoff? To give a 
mathematically satisfactory answer to this question is quite difficult; the 
interested reader should see the references in Figure \ref{fig:errortable}. It turns out,
however, that by returning to our medical analogy and studying
how a doctor selects controls for Mr.\ Payne, we obtain a much more
tractable problem which holds the same lessons as the original. Readers
satisfied with the heuristics given above may skip the following section, which is inevitably more technical. 

\subsection{ A Cost Benefit Analysis of Matching}

\label{sec:analogy}

The doctor wants to know $y$, Mr.\ Payne's response if he were assigned the
newly approved treatment $B$. Just as we estimate $\Delta$ by $\bar{\Delta
}^{\prime}$ , the average error over control problems, so the doctor estimates
$y$ using $\bar{y}^{\prime}$, the average response to treatment $B$
over some set of control patients. We can hopefully improve upon this initial
estimate by exploiting the fact that some patients are more relevant for
Mr.\ Payne than others. But
achieving higher relevance comes with the obvious cost of a reduction in the
number of available controls. To what degree then should controls match the
individual of interest? 

\textbf{The ideal case of infinitely many matches. }We begin our analysis by
removing one side of the tradeoff: the constraint that our clinical trial
enrolls only a finite number of patients. Let $\Omega$ denote an
infinite set of control patients and $\Omega_{\vec{a}}$ the subset with
background variables $\vec{x}^{\prime}=\vec{a}$. In particular, $\Omega
_{\vec{x}}$ consists of those patients who match Mr.\ Payne's observed
characteristics completely. We show that the average treatment response for
patients in $\Omega_{\vec{x}},$ denoted $\mu_{\vec{x}}$, is a better estimate
of Mr.\ Payne's response $y$ than the average response for patients in $\Omega
$, denoted $\mu$. Our criterion will be the \textit{mean-squared error} (MSE). Ideally we want to compare $\left(  y-\mu_{\vec{x}}\right)  ^{2}$ and $\left(
y-\mu\right)  ^{2}$, but because they are unknown, we estimate them by their corresponding
averages over $\Omega$ (or $\Omega_{\vec{x}}$).
That is, for each  $\left(  \vec{x}^{\prime
},y^{\prime}\right)  $ in $\Omega$, we compute the accuracy of $\mu$ 
 in predicting $y^{\prime}$; we then repeat this assessment for $\mu_{\vec{x}^{\prime}}$.  Because $\left(y^{\prime}-\mu\right)^{2}=\left(  y^{\prime}-\mu_{\vec{x}^{\prime}}\right)^{2}+\left(\mu_{\vec{x}^{\prime}}-\mu\right)^{2}+
2\left(  y-\mu_{\vec{x}^{\prime}}\right)\left(\mu_{\vec{x}^{\prime}}-\mu\right),$
and the last term is zero when averaged over $\Omega$, we have
\begin{equation}\label{eq:anova}
\text{Ave}_{\Omega}\left[\left(y-\mu\right)^{2}\right] - \text{Ave}_{\Omega}\left[\left(  y-\mu_{\vec{x}^{\prime}}\right)^{2}\right]
= \text{Ave}_{\Omega} \left[\left(\mu_{\vec{x}^{\prime}}-\mu\right)^{2}\right].
\end{equation}
Here, Ave$_{\Omega}\left[  \cdot\right]$ is the average of the bracketed expression across controls
in $\Omega$. 
We see that the error from using $\mu_{\vec{x}}$ to predict $y$ is (on average) no worse than that from using $\mu$. 
And the expected improvement---$\text{Ave}_{\Omega}\left[ (\mu-\mu_{\vec{x}^{\prime}})^2 \right]$---is precisely the average (squared) difference between 
the more relevant and less relevant estimates of $y^{\prime}$. 

\textbf{Finite-sample complications.} Real clinical trials have only a finite
number of patients. Let $\hat{\Omega}$ denote this finite set and $\hat
{\Omega}_{\vec{a}}$ the subset of patients with $\vec{x}^{\prime}=\vec{a}$;
analogously, $\hat{\mu}$ and $\hat{\mu}_{\vec{a}}$ denote the average patient
response in $\hat{\Omega}$ and $\hat{\Omega}_{\vec{a}}$ respectively. In the
case of unlimited controls,  on average $\mu_{\vec{x}}$ is a better predictor
of $y$ than $\mu$; with only a finite number of controls, it is now 
possible for $\hat{\mu}$ to outperform $\hat{\mu}_{\vec{x}}$. This is because
we are implicitly engaging in a two-stage estimation process---(a) estimate
$y$ by $\mu$ (or $\mu_{\vec{x}}$) and (b) estimate $\mu$ by $\hat{\mu}$ (or $\mu_{\vec{x}}$ by $\hat{\mu}_{\vec{x}}$). Since few individuals match Mr. Payne's background
completely, it is harder to estimate $\mu_{\vec{x}}$ than $\mu.$ This increase
in estimation error can wipe out benefits from matching. 

\textbf{Finding the sweet spot}. When the cost of estimating $\mu_{\vec{x}}$
by $\hat{\mu}_{\vec{x}}$ is too high, we can compromise by using controls who
match Mr. Payne with respect to some, but not all, variables. Let $\vec
{x}_{(0)}^{\prime},\vec{x}_{(1)}^{\prime},...,\vec{x}_{(R)}^{\prime}$ denote
sequentially expanding\ collections  of background variables: for example, $\vec
{x}_{(1)}^{\prime}$ could be age, $\vec{x}_{(2)}^{\prime}$ age \textit{and}
blood pressure, etc. We use $\vec{x}_{(0)}^{\prime}$ to denote no matching and
$\vec{x}^{\prime}=\vec{x}_{(R)}^{\prime}$ to denote matching on all the
measured variables. A natural question then is: \textit{given such a
sequence}, for what value of $r$ between $0$ and $R$, does the incremental increase in estimation error
exceed the incremental gain from matching? (An even harder question is how to
order the sequence so that the \textquotedblleft most
important\textquotedblright\ predictors appear first; see
\citealt{meng2014trio} for a discussion.)

To answer this question, we want to compare $(  y-\hat{\mu}_{\vec{x}_{\left(  r\right)  }})
^{2}$ versus $(y-\hat{\mu}_{\vec{x}_{(r+1)}})^{2}$, i.e.,  matching on $r$ versus
$r+1$ background variables. Again, these two prediction errors are unknown, so
we will estimate them by their corresponding averages over control patients,
say in $\Omega$. The average
prediction error for $\hat{\mu}_{\vec{x}_{\left(  r\right)  }^{\prime}}$ minus
the average prediction error for $\hat{\mu}_{\vec{x}_{\left(  r+1\right)
}^{\prime}}$ decomposes as follows:
\begin{equation}
\label{risk3}
\underbrace{\text{Ave}_{\Omega}\left[  (\mu_{\vec{x}_{(r)}^{\prime}}-\mu
_{\vec{x}_{\left(  r+1\right)  }^{\prime}})^{2}\right]  }_{\text{gain in
relevance}} \qquad -\qquad \underbrace{\text{Ave}_{\Omega}\left[  (\hat{\mu}_{\vec
{x}_{\left(  r+1\right)  }^{\prime}}-\mu_{\vec{x}_{(r+1)}^{\prime}}%
)^{2}\right]  -\text{Ave}_{\Omega}\left[  (\hat{\mu}_{\vec{x}_{(r)}^{\prime}%
}-\mu_{\vec{x}_{(r)}^{\prime}})^{2}\right]  }_{\text{loss in robustness}%
}.%
\end{equation}
The second term compares how
difficult it is to estimate $\mu_{\vec{x}_{\left(  r+1\right)  }^{\prime}}$
versus $\mu_{\vec{x}_{\left(  r\right)  }^{\prime}}$---since it is harder to
find controls which match on $r+1$ variables as compared to $r$, $\mu
_{\vec{x}_{\left(  r+1\right)  }}$ is usually harder to estimate. So, the
second term is usually positive; it represents the loss in robustness from an
additional step of matching. The first term is analogous to the RHS of equation (\ref{eq:anova}), i.e., it represents the benefit of matching. Therefore, expression (\ref{risk3}) lays out mathematically the relevance robustness tradeoff.

\bigskip
\textbf{How this all relates to statistical inference}. It may be difficult at
first to see why the lessons above transfer over to statistical inference.
After all, what limits the doctor's ability to match is the \textit{finite}
nature of her clinical trial, whereas statisticians can simulate as many
control problems $\left(  D^{\prime},\theta^{\prime}\right)  $ as they want, at least in principle.
The key is this: having a finite clinical trial means that if we rerun the
trial by sampling $n$ new patients, our inferences would change. That is, $\hat{\Omega
}$ is unstable. The ideal trial $\Omega$---by virtue of enrolling \textquotedblleft
everyone\textquotedblright---is free of the idiosyncrasies of any sample; it
is stable.  In the statistical context, the set of control problems, $\hat \Omega$, is also 
``unstable" because we cannot specify the patterns $\hat f_{\theta}$ and $\hat f_{\varepsilon}$
with perfect certainty. In both contexts, the instability of $\hat \Omega$ is
magnified when considering subsets of controls---in this way, 
matching erodes robustness.

\subsection{Matching: The Doctor Does It, So Should the Statistician}

\label{sec:relevance}

The more closely a clinical trial patient matches Mr.\ Payne's observed characteristics, $\vec{x}$, the more relevant the patient is. Similarly, relevant control problems,
$\left(  D^{\prime},\theta^{\prime}\right),$ will match our target problem,
$\left(  D,\theta\right)  $, with respect to its ``observed characteristics": $D$.
Complete matches are controls with $D^{\prime}=D.$ This correspondence between $\vec{x}$ and $D$ makes intuitive sense: if
the data in a control problem fail to mimic important features of our actual data, evaluating a
procedure on the former tells us little, or may even mislead us, about its
performance on the latter. In the words of \citet{kiefer1976,kiefer1977}, we
need to distinguish those problems with \textquotedblleft
lucky\textquotedblright\ data from those with \textquotedblleft
unlucky\textquotedblright\ data, as illustrated below. \bigskip

\bigskip

\textbf{Example 2 (Data Precision; \citealt{cox1958})}. Suppose we have $n$
independent measurements on the weight of an object. The sample size, $n$, is
a feature of the data (an often forgotten fact); since problems with very
different sample sizes are not comparable, we want our controls to have a
matching sample size of $n$. Suppose we also know that the task of taking the
$i$th measurement was randomly assigned to one of two labs; the first lab
produces very precise estimates, the second lab noisier ones. By chance, all
but two measurements in \textit{our }sample came from Lab 2. The lab
assignments are also an important feature of the observed data. If a control
problem has data where the majority of measurements originate from the more
precise Lab 1, an estimator's effectiveness on the control may be an overly
optimistic assessment of its error when applied to our data. Hence, we
should also match on lab assignment.

Sample size and lab assignment help determine the \textit{precision }of our
data; in other contexts, other features of the data may help determine
precision. For example, suppose we are analyzing data from a randomized
clinical trial comparing treatments $A$ and $B$. By chance, 80\% of females
received treatment $A$; this imbalance makes it harder to disentangle the
treatment effect from any gender effect. If control datasets do not preserve
this gender imbalance, the estimate of the treatment effect would appear to be
much more precise than it actually is, causing us to be overconfident. Hence
\citet{cox1982} and \citet{rosenbaum1984a} suggest keeping only those control
problems where the covariate balance is sufficiently similar to the observed
balance. Both examples above teach us that the precision of control data
should match that of our actual data. This seemingly obvious principle
is violated surprisingly often in practice; see \citet{fraser2004} for discussion.

\bigskip

In problems involving \textit{independent} measurements, matching on the sample size is
standard practice. The qualitative effect of matching---larger sample sizes
imply greater precision, smaller ones less precision---is robust to our
initial selection of control problems, $\hat{\Omega}$, i.e., robust to model
misspecification. As Example 3 below shows, this robustness property is not shared
by all features of the data. The decision whether to match on these other
features is much more controversial. In some situations, even matching on the
sample size can lead to non-robust inferences \citep[see][]{rosenbaum1984b}.

\bigskip

\textbf{Example 3}. Let $y$ be a single measurement for the weight, $\theta$,
of an object; the natural estimator of $\theta$ is $y$ itself. To assess its accuracy, we simulate controls as follows: $y^{\prime}=\theta^{\prime
}+\varepsilon^{\prime}$, where the measurement errors $\varepsilon^{\prime}$
vary according to a distribution $f_{\varepsilon}$ (supplied by the scale
manufacturer) and we set $\theta^{\prime}=c$. For each choice of the
simulation parameter $c$, we obtain a different set of control problems.
However, our estimator's average error $\left\vert y^{\prime}-\theta^{\prime}\right\vert $
over $\hat{\Omega}$ turns out to be independent of $c$. This robustness vanishes
when we match on the measurement itself; the average error over controls
with $y^{\prime}=y$ is $\left\vert y-c\right\vert $ , which can range anywhere
from $0$ to $\infty$ depending on $c$. Intuitively, the value of the
measurement $y$ tells us nothing  about its precision \textit{unless} we
have some external information about $\theta$. If we do have this information, say $\theta < 10$, then
the value of $y$ can be very informative about its precision, e.g., $y>10$ would be an imprecise measurement. 

In order to justify matching on $y$ in this problem (as is done in Bayesian inference),
we must know something (reliable) about $\theta$. When we do not, can we somehow specify an ``objective/non-informative" choice of $\hat f_{\theta}$ and obtain a sensible inference from matching on $y$? This quest underlies
 objective Bayesian analyses \citep[see][]{kass1996,berger2006}. Such analyses usually choose $\hat f_{\theta}$
with the goal of harmonizing the inference from matching on $y$ with the inference from not matching on $y$.   
Thus, while we nominally obtain an ``individualized" inference, the matching has no value added. 
In this example, there is no way to
obtain \textit{meaningful} ``individualization" without genuine
information about $\theta$.

\bigskip

The features of the data described in Example 2 are examples of
\textit{ancillary }statistics \citep[see][]{fraser2004,ghosh2010}; the idea of
matching on ancillaries dates back to \citet{fisher1925,fisher1934}. The name,
ancillary, comes from the fact\ that such features of the data can be
simulated \textit{without }specifying $\hat{f}_{\theta}$ (this gives an
intuitive way to see why matching on such features does not erode the robustness
of our inference). In contrast,
the measurement $y^{\prime}$ in Example 3 is not ancillary---to simulate
$y^{\prime}$, we first simulate $\theta^{\prime}$. Ancillary statistics tell
us something about the precision of our data even when we know nothing about
$\theta$. Non-ancillary statistics also carry information on the precision of
our data, but the effective use of this information often
requires prior knowledge about $\theta$. Many features of the data lie in
between the extremes of Example 2 and Example 3, i.e., the information
they contain about precision is partially  sensitive to $\hat
{f}_{\theta}$ (see Examples 4 and 5). The relevance robustness tradeoff
becomes much harder to navigate in such cases. 

\bigskip

\textbf{Example 4 (Accounting for Selection Bias)}. Using data from $n$
patients, we wish to identify genetic markers which influence an individual's
cholesterol level. We estimated each marker's effect by looking at the
difference in average cholesterol between those with and without the marker;
denote our estimates by $\hat{\theta}_{m}$ (for $m=1,...,M$). We also computed
a p-value for each marker assessing whether $\hat{\theta}_{m}$ is
significantly different from $0$. Only markers with p-value below
some threshold are selected for further study. But how accurate are the
estimates $\hat{\theta}_{m}$ for the \textit{selected }markers?

Let us focus on marker 1. The key is to realize that there are two types of
datasets---those in which marker 1 is selected and those in which it is
not---and that the accuracy of $\hat{\theta}_{m}$ differs (often dramatically)
between the two types. To see this, note that two factors help determine
whether a marker is selected: (i) the magnitude of the marker's impact on
cholesterol and (ii) luck/chance. Luck in this case means that the magnitude
of our estimate $\vert \hat{\theta}_{m}\vert $ is upward biased
compared to the true magnitude (i.e., luck helps marker $m$ get selected). When
we perform selection, we are targeting not only markers with large effects,
but also markers for which the current dataset is \textquotedblleft
lucky\textquotedblright. Our original estimates for the selected markers
likely overstate their impact. Bias induced by selection is commonly referred
to as \textquotedblleft winner's curse\textquotedblright%
\ \citep{ioannidis2001}. We want to remove this curse.

Suppose we have some model for simulating control problems $\left(  D^{\prime
},\theta^{\prime}\right)  $. The reasoning above tells us that we should
restrict attention to controls where the same markers are selected as in our
actual study; without this matching step, our inference ignores selection
bias. After matching, we can compute the bias of $\hat{\theta}_{m}^{\prime}$
over the relevant controls; this acts as an estimate of the actual bias of
$\hat{\theta}_{m}$. However, there is a problem: the bias of $\hat{\theta}%
_{m}^{\prime}$ over the relevant controls is sensitive to how we simulate
$\theta^{\prime}$ (i.e., sensitive to $\hat{f}_{\theta
}$). If a marker's true effect size, $\left\vert \theta_{m}^{\prime
}\right\vert $, is large, it needs very little luck to be selected, so the
bias of $\hat{\theta}_{m}^{\prime}$ will be small; the converse holds when
$\left\vert \theta_{m}^{\prime}\right\vert $ is small. To effectively estimate
the winner's curse by matching, we need a reliable specification of $\hat
{f}_{\theta}$; fortunately this is often possible when the number of markers
is large \citep[see][]{efron2010}.

\bigskip

\begin{remark}
The above example is a variation on the statistical practice of \textit{model
selection}. When fitting a model of outcome on predictors, a selection step
may be introduced to screen out unimportant predictors, e.g., using criterion
such as AIC and BIC \citep[see][]{burnham2002}. This practice is becoming
increasingly common because modern applications often involve so-called
\textquotedblleft large-\textit{p} small-\textit{n}" data where the number of
predictors exceeds the number of observations. One may then wish to estimate
and build confidence intervals for the parameters in the selected model.
Ignoring the selection bias will lead to seriously flawed inferences \citep[see][]{leeb2005}.
\end{remark}

\bigskip

\textbf{Example 5 (The Strength of Evidence)}. This example comes from
\citet{dempster1997} and is extended using ideas in \citet{berger2003}. We
want to test a null hypothesis, $H_{0}:\theta=0$, against an alternative,
$H_{1}:\theta=1$. Suppose we observe a p-value $p=0.049$, and reject the null
hypothesis according to a pre-determined testing procedure: reject $H_{0}$ if
$p\leq0.05$. We want to assess the probability that we have made a false
rejection. So we evaluate our testing protocol over some relevant subset of
control problems $(p^{\prime},\theta^{\prime})$ in $\hat{\Omega}$ (for
simplicity, we assume here that $p^{\prime}$ preserves all the information in
the data). The question is: which subset?

\begin{enumerate}
\item[\textbf{Case 1}.] \textit{Use All Control Problems in }$\hat{\Omega}$:
p-values are constructed to follow a uniform distribution when $H_{0}%
:\theta^{\prime}=0$ is true. Hence, our error rate (proportion of times
$p^{\prime}\leq0.05$) over control problems with $\theta^{\prime}=0$ is 5\%;
this is the Type I error of our test. Suppose the p-value over problems with
$\theta^{\prime}=1$ follows a Beta$(0.02,1.35)$ distribution (see Figure
\ref{fig:pvalue}a)---that is, we are more likely to see very small p-values
when $H_{1}$ is true. Under this Beta distribution, 5\% of datasets with $\theta^{\prime}=1$ have
$p^{\prime}>0.05$ so our error rate under $H_{1}$ (Type II error) is also 5\%.
Thus, \textit{regardless of} $\hat{f}_{\theta}$, the error rate of our test
over all problems in $\hat{\Omega}$ is 5\%.
\end{enumerate}

\begin{figure}[ptbh]
\begin{center}
\includegraphics[width=4in]{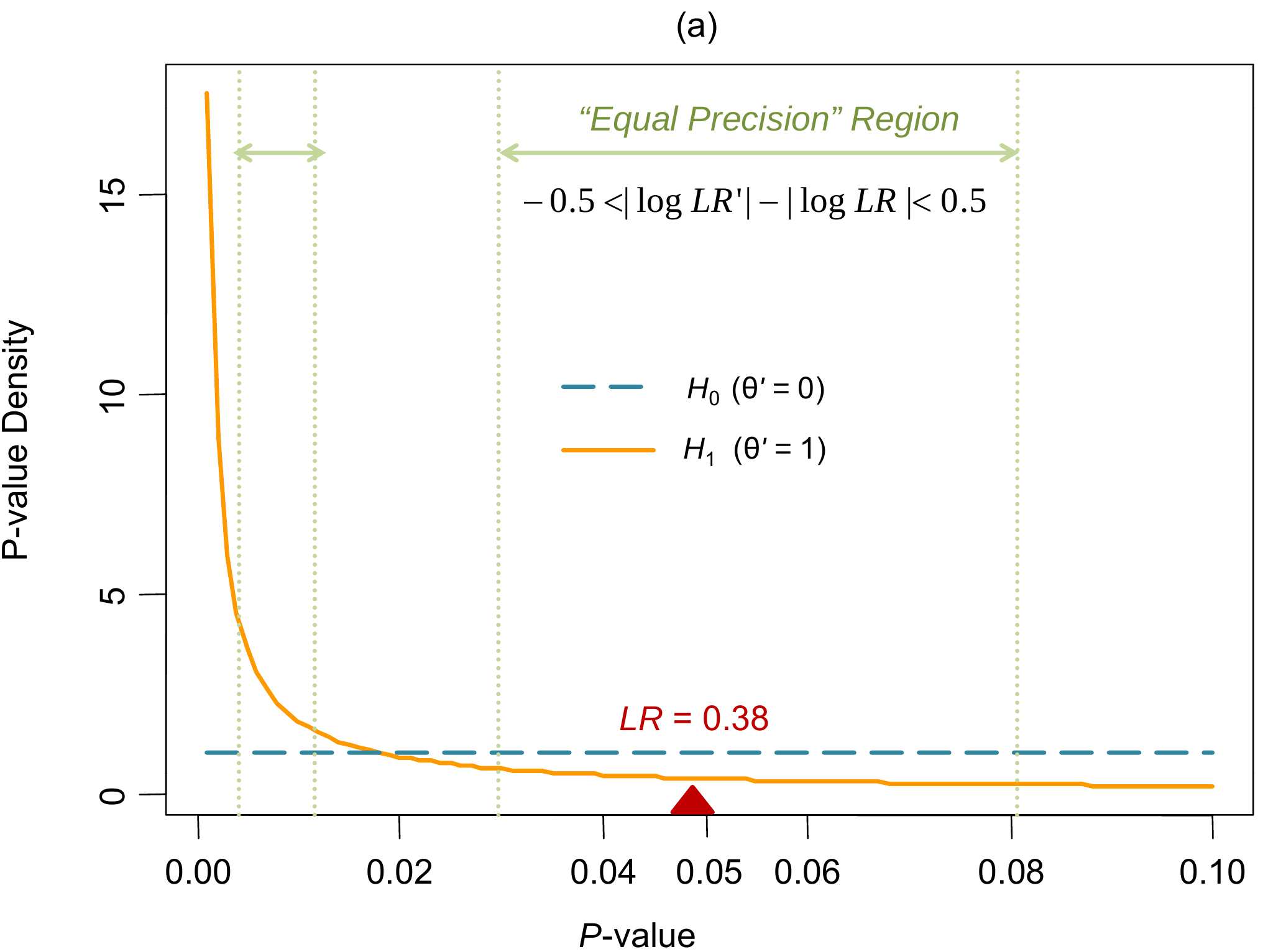} \vskip 0.2in
\includegraphics[width=4in]{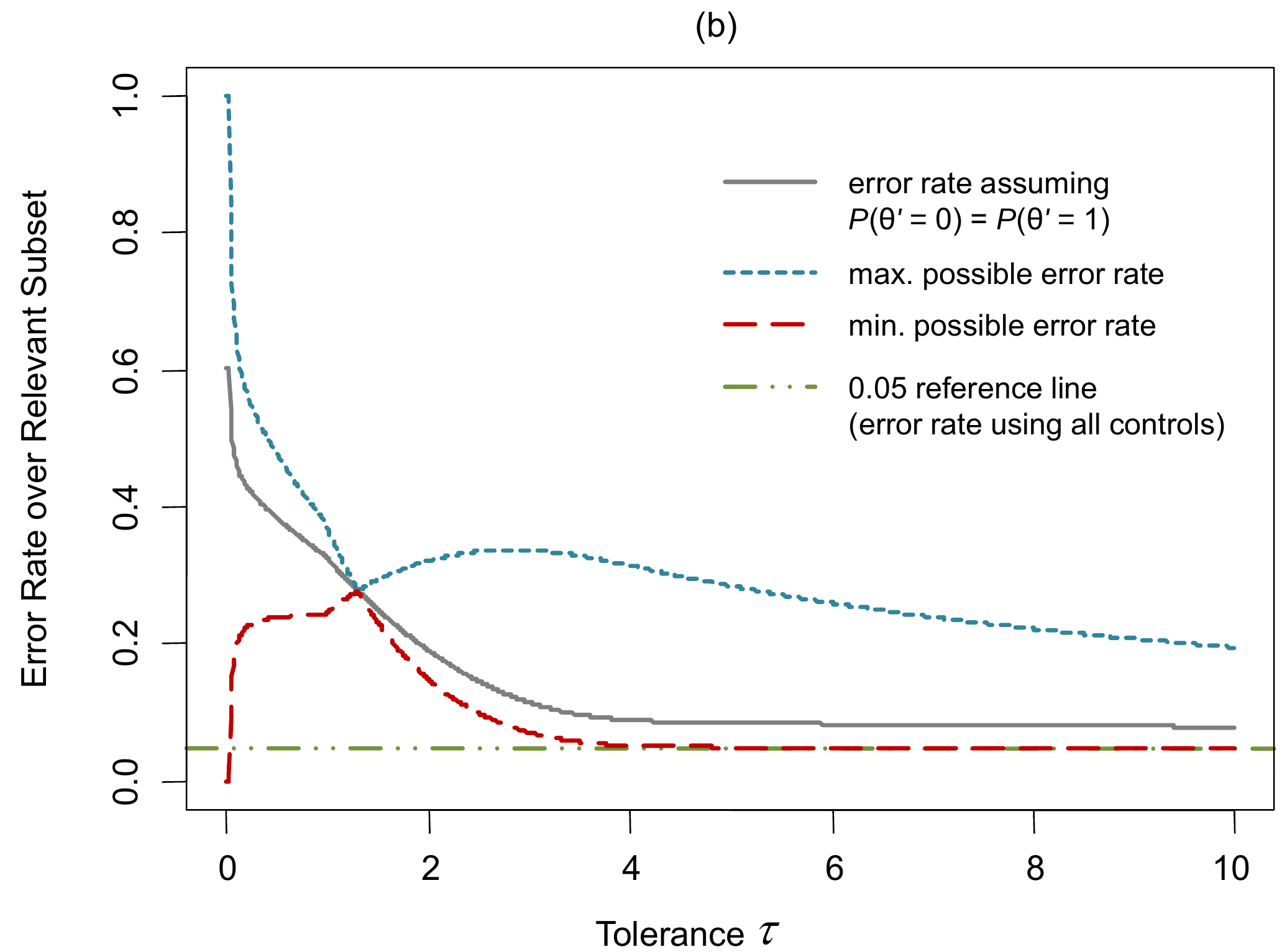}
\end{center}
\caption{Plot (a) depicts the densities of the p-value distribution when
$H_{0}$ or $H_{1}$ are true. The density under $H_{1}$ is higher for small
$p,$ i.e., small p-values favor $H_{1}$. The red triangle marks the position
of the observed p-value, whose density under $H_{0}$ exceeds its density under
$H_{1}$ (the evidence favors $H_{0}$). The dashed vertical lines demarcate an ``equal precision region" (this is a disjoint region): only controls with an observed p-value,  $p^{\prime}$, lying in this region are considered sufficiently relevant for the actual problem. 
Here, ``sufficiently relevant" is specified through the tolerance $\tau=0.5$ . The
solid gray line in (b) plots the error rate of our test over $\hat{\Omega
}_{\text{rele}}$ for various values of $\tau$ (when $\hat f_{\theta}$ assigns equal
probability to $H_{0}$ and $H_{1}$). The dashed lines show the maximal and
minimal possible error rates over $\hat \Omega_{\text{rele}}$ when $\hat{f}_{\theta}$ is 
allowed to be arbitrary.
Sensitivity to $\hat{f}_{\theta}$ increases as $\tau\rightarrow0$, that is, as we
increase the relevance.}%
\label{fig:pvalue}%
\end{figure}

Despite its appealing robustness, we notice some unsettling facts about the
above inference. 

\begin{itemize}
\item If we had observed $p=10^{-8}$, our error assessment (using $\hat
{\Omega}$) would still be 5\%. Yet the chance of false rejection is
intuitively much smaller when $p=10^{-8}$ than when $p=0.049$.

\item In Figure \ref{fig:pvalue}a, the density of the p-value distribution at
$p=0.049$ is higher when $H_{0}$ is true than when $H_{1}$ is true. The ratio
of the density under $H_{1}$ to that under $H_{0}$ is $LR=0.38$. We call this
the \textit{likelihood ratio}. A $LR<1$ says that the data supports $H_{0}$
more than $H_{1}$! Given this, a 5\% error rate seems an inaccurate assessment
of our uncertainty.
\end{itemize}

These incongruities have a simple explanation. The error rate over
$\hat{\Omega}$, measures the \textit{average accuracy} of a control problem.
But a rejection based on $p^{\prime}=10^{-8}$ is much more convincing than one
based on $p^{\prime}=0.049$. So the average accuracy will underestimate the
actual accuracy when $p=10^{-8}$ and overestimate it when $p=0.049$.

To fix this, we need to match on some measure of precision. The likelihood
ratio seems to capture precision reasonably well, but  we will make one small
modification to it. Note that a testing problem with $LR^{\prime}=\frac{1}{2}$
(evidence favors $H_{0}$ by factor of $2$) seems as difficult as one with
$LR^{\prime}=2$ (evidence favors $H_{1}$ by factor of $2$). This symmetry
motivates us to define as \textit{relevant} those controls for which
either\textit{ }$LR^{\prime}\approx LR$ or\textit{ }$LR^{\prime}\approx
LR^{-1}$; more compactly, controls such that $\left\vert \log LR^{\prime
}|-|\log LR\right\vert \approx0$ \citep[see][for an argument which leads to matching on ancillary features]{berger1994}.

\begin{enumerate}
\item[\textbf{Case 2}.] $\hat{\Omega}_{\text{rele}}$ \textit{contains those
problems }$(p^{\prime},\theta^{\prime})$ \textit{in }$\hat{\Omega}$
\textit{such that} $-\tau\leq\left\vert \log LR^{\prime}\right\vert -|\log
LR|\leq \tau$. The tolerance $\tau$ measures the degree to which we require our
controls to match our actual problem; for $\tau=0.5$, $\hat{\Omega}_{\text{rele}}$ includes only controls with observed 
p-value, $p^{\prime}$, in the ``equal precision region'' (see Figure \ref{fig:pvalue}a). When $\tau=0$, $LR^{\prime}=0.38,$ or
$1/{0.38}$---our control has the same precision as the actual problem. When
$\tau=\infty$, we use all controls in $\hat{\Omega}$; this is equivalent to Case 1. 
\end{enumerate}

Unlike Case 1, the error rate of our test over $\hat{\Omega}_{\text{rele}}$
depends on our specification of $\hat{f}_{\theta}$; Figure \ref{fig:pvalue}b
plots the maximal and minimal error rates (dashed lines) that can be achieved by varying
$\hat{f}_{\theta}$ (this helps us assess how sensitive our inference is).
Figure \ref{fig:pvalue}b also plots the error rate over $\hat{\Omega
}_{\text{rele}}$ when $\hat{f}_{\theta}$ assigns equal probability to $H_{0}$
and $H_{1}$ (a popular choice).

As we increase the relevance ($\tau \rightarrow0$), inferences drawn from
$\hat{\Omega}_{\text{rele}}$ becomes less and less robust: the error rate
ranges anywhere from $0$ to $1$ depending on how we specify $\hat{f}_{\theta}%
$. However, for almost all reasonable choices of $\tau$, the range of plausible
error rates lies above 5\%---the error rate when we do not match. This
suggests that, at the least, the qualitative impact of matching---to correct
the downward bias of our initial error estimate---is quite robust even if the degree of adjustment is not.
Intuitively, the more \textit{heterogeneous} the problems in $\hat{\Omega}$,
the more important it is to match. In this example, the performance of our
testing procedure is sufficiently different when $p^{\prime}=10^{-8}$ than
when $p^{\prime}=0.049$ to justify some degree of matching---of course,
deciding the exact degree is a much harder problem! Interestingly,
\citet[Example 4b]{berger1988} give an even more extreme form of this example
where $\hat{\Omega}$ is so heterogeneous that complete matching is warranted. 

\bigskip

In summary, Example 2 is the poster child for matching (relevance), Example 3
the burning beacon of its potential dangers; Examples 4 and 5 fall somewhere in between. In choosing the right relevance robustness tradeoff for a problem, it is helpful to think about how it maps in terms of these canonical examples. 

\section{Robustness Through Symmetry and Invariance}

\label{sec:robustness}

To increase the relevance of controls in our approximating population, we
match on features of the data. Is there any way to increase robustness? More
precisely, can we design procedures which perform well regardless of $\hat{f}_{\theta}$
? Consider Example 5: the average error of our test over $\hat{\Omega}$
was 5\% for any choice of $\hat{f}_{\theta}$. This
robustness derives from the \textit{symmetry} or \textit{invariance} of the
procedure. The accuracy of our
hypothesis test (over $\hat \Omega$) is invariant to whether the null or alternative hypothesis
is true. As the following example demonstrates, \textit{conditioning}
(the technical term for matching) on non-ancillary features of the data
breaks the symmetry of a procedure. The implication is that it is hard
to construct procedures which are symmetric not only over $\hat \Omega$ but also over \textit{subsets} of $\hat \Omega$.
The intuition for this phenomenon is simple: it's hard designing medicine that works equally well
on all individuals regardless of background.  

\bigskip

\textbf{Example 6 (Too High a Price for Relevance)}. A doctor wants to predict
your disease status: sick ($\theta=1$) or healthy ($\theta=0$)? A test for the
disease yields a positive result, $y=1$. The company producing the test tells
the doctor that, in a clinical trial $\hat{\Omega}$, sick
individuals tested positive with 90\% probability while healthy ones tested
positive with 10\% probability. To decide whether to initiate treatment, the
doctor asks for the probability that \textit{your test result }is wrong.

Observe that the test is wrong 10\% of the time for sick individuals (false
negative) and 10\% of the time for healthy individuals (false positive), i.e.,
the test's accuracy is invariant to the patient's underlying disease status. Hence,
the test's error rate with respect to the clinical trial population,
$\hat{\Omega}$, is 10\%. This assessment does not depend on knowing the proportion
of sick individuals ($\theta^{\prime} =1$) in the clinical trial. This proportion, which
is potentially unknown to the doctor, acts as $\hat f_{\theta}$ in our example.

Not all clinical trial patients are
relevant \textit{for you}: your test came back positive while the clinical trial
contains individuals with negative results. What the doctor actually seeks is
the accuracy of the test among the subset, $\hat{\Omega}_{\text{rele}}$,  of individuals \textit{who tested positive}. 
Among this subset of patients, the accuracy of the test is no longer invariant to the patient's underlying disease status: 
the test is 100\% accurate for the sick and 0\%
accurate for the healthy. The test is symmetric over $\hat \Omega$ but not over $\hat{\Omega}_{\text{rele}}$.
Due to this loss of invariance, the test's accuracy among
patients in $\hat{\Omega}_{\text{rele}}$ becomes sensitive to the proportion of sick individuals. 
In fact, the error rate can range anywhere from 0\% to 100\%!

\bigskip

Example 6 shows how ``symmetry" allows us to make inferences that are robust to misspecification of $\hat f_{\theta}$
\citep[see][for an application of this idea in the case of randomized experiments]{rubin1978}. The price, of course, is that
we need to limit the degree to which we ``individualize" our inference---procedures which act symmetrically over a set of controls often behave
asymmetrically over its subsets. The pressing question now is how to construct symmetric procedures. It turns out that
the study of symmetry is best initiated by studying \textit{a}symmetry.

\subsection{Asymmetric Procedures and the Plug-in Principle}

\label{sec:pivots}

A procedure may behave symmetrically in one statistical problem and asymmetrically in another. 
 To illustrate this point, we consider a confidence interval which behaves asymmetrically because it ``violates"
 the problem's natural symmetries.

\bigskip

\textbf{Example 7 (An Asymmetric Procedure). }For quality control purposes, a
company records failure times $y_{i}$ for $n$ randomly selected batteries. The
goal is to find a lower bound on the average time to battery failure, $\theta
$. Clearly, we cannot bound $\theta$ deterministically, except for the useless bound zero. To create a
probabilistic lower bound we begin with an estimate of $\theta$, e.g., the
sample mean $\bar{y}_{n}$. A seemingly innocuous construction is to then
choose a buffer $b$ so that the lower bound $\bar{y}_{n}^{\prime}-b$ lies
below $\theta^{\prime}$ for a desired fraction of controls; here $D^{\prime}=\left\{  y_{i}^{\prime
}\right\}  _{i=1}^{n}$. 

To simulate controls, the company uses a multiplicative model for the failure
times: $y_{i}^{\prime}=\theta^{\prime}\varepsilon_{i}^{\prime}$, where
the mean failure time $\theta^{\prime}$ is simulated from some
 $\hat{f}_{\theta}$, and the $i$th battery's deviation from the mean, $\varepsilon_{i}^{\prime}>0$,
is independently drawn from a distribution $\hat{f}_{\varepsilon}$
with mean 1. The difficulty lies in specifying a reasonable
$\hat{f}_{\theta}$, which requires prior knowledge of the average failure time.
For our inference to be robust, we want to choose $b$ such that $\bar{y}%
_{n}^{\prime}-b$ is a good lower bound regardless of $\hat{f}_{\theta}$. Does
a non-trivial choice exist?

The answer is no. We have purposely cooked up an \textquotedblleft
incompatibility\textquotedblright\ in our example: our bound $\bar{y}_{n}-b$
implicitly assumes an additive structure for battery heterogeneity, but our
controls are generated from a multiplicative model. This incompatibility makes
the performance of our bound ultra-sensitive to any choice of $\hat{f}%
_{\theta}$. Since $\bar{y}_{n}^{\prime}=\theta^{\prime}\bar{\varepsilon}%
_{n}^{\prime}$, the lower bound succeeds ($\theta^{\prime}>\bar{y}_{n}%
^{\prime}-b$) for a control problem if and only if $\bar{\varepsilon}%
_{n}^{\prime}<1+b/\theta^{\prime}$. We can think of $1+b/\theta^{\prime}$ as
an upper bound for $\bar{\varepsilon}_{n}^{\prime}$ with buffer $b/\theta
^{\prime}$, which gets arbitrarily  close to zero as we increase  $\theta^{\prime}$. The lower bound's
effectiveness therefore is asymmetric, i.e., it depends on the particular value of
$\theta^{\prime}$,  yielding no useful lower bound that  will work well for all $\hat{f}_{\theta}$. 
\bigskip

Example 7 shows that we cannot assess the effectiveness of an asymmetric
procedure in a manner robust to $\hat{f}_{\theta}$. Hence, there is little
reliable information for deciding whether to use that procedure to analyze
our dataset---if we use it, we are at the mercy of luck or unluck. Yet asymmetric
statistical procedures are often unavoidable simply because non-trivial
symmetric procedures do not exist. A crude but convenient way
\citep[see][]{efron1998} to proceed is to \textquotedblleft
plug-in\textquotedblright\ an estimate of $\theta$ when generating controls,
i.e., set $\theta^{\prime}=\hat{\theta}$ where $\hat{\theta}$ is our estimate.
This is equivalent to specifying $\hat{f}_{\theta}$ as a point mass at
$\hat{\theta}$; the \textit{parametric bootstrap}
\citep[]{efron1994,davison1997} and \textit{Empirical Bayes} (see Section
\ref{sec:diagonal}) are common applications of this strategy. The plug-in
approach yokes our uncertainty assessment to an initial estimate of $\theta$.
If a procedure's effectiveness is highly sensitive to the value of $\theta$
(i.e., highly asymmetric), the resulting inference may be worthless for even
small errors in our initial estimate. That is, our error assessment will be
least reliable precisely when our estimation error is greatest---a most
unattractive quality! The plug-in approach therefore works best when our
procedure is nearly symmetric (this point adds additional motivation for
understanding how to create symmetric or nearly symmetric procedures). Despite its
dangers, plugging in is popular in practice because of its ease of implementation, but it should really be treated as a weapon of desperation.

\subsection{Symmetry/Invariance  Creation Via Pivots and Minimaxity}

\label{sec:minimax}

\qquad We now discuss two strategies for constructing procedures that are
symmetric in $\theta$ or approximately so. Such procedures have
performance guarantees that are robust to the user's specification of $\hat
{f}_{\theta}$. We may desire robustness to $\hat{f}_{\varepsilon}$ as
well as $\hat{f}_{\theta}$. Extensions of the strategies below can often lead
to procedures that are symmetric in both $\varepsilon$ \textit{and} $\theta$;
the interested reader can follow the path to randomization inference (see
Figure \ref{fig:inftable2}). Of course, the cost of such robustness is a
further drop in relevance \citep[see][]{basu1980}.

\bigskip

\qquad\textbf{Creating Symmetry/Invariance  via Pivots}. As motivation, consider this
observation from Example 7: the representation $\bar{y}_{n}^{\prime}%
=\theta^{\prime}\cdot\bar{\varepsilon}_{n}^{\prime}$ induces a duality between
$\theta^{\prime}$ and $\bar{\varepsilon}_{n}^{\prime}$---a lower bound
($\bar{y}_{n}^{\prime}-b$) for $\theta^{\prime}$ induces an upper bound
($1+b/\theta^{\prime}$) for $\bar{\varepsilon}_{n}^{\prime}$ and vice versa.
The fact that the effectiveness of this procedure depends on $\theta^{\prime}$
is self-evident by considering the upper bound for $\bar{\varepsilon}%
_{n}^{\prime}$ (which is small if $\theta^{\prime}$ is large). Intuitively, we
can \textquotedblleft symmetrize\textquotedblright\ our procedure by using an
upper bound for $\bar{\varepsilon}_{n}^{\prime}$ that is \textit{independent
}of $\theta^{\prime}$. The performance of our procedure will then be \textit{invariant} to the choice of $\hat f_\theta$.\ For example, we can choose a constant $c$ such that
$\bar{\varepsilon}_{n}^{\prime}<c$ for 95\% of the controls in $\hat{\Omega}$.
By duality, we automatically obtain $\theta^{\prime}>\bar{y}_{n}^{\prime}/c$
for 95\% of the controls in $\hat{\Omega}$. Note that:

\begin{itemize}
\item The distribution of $\bar{\varepsilon}_{n}^{\prime}$ depends only on
$\hat{f}_{\varepsilon}$ so our choice of $c$ does not depend on $\hat
{f}_{\theta}$. As a result, the guarantee---$\theta^{\prime}>\bar{y}%
_{n}^{\prime}/c$ for 95\% of controls in $\hat{\Omega}$---holds for any choice
of $\hat{f}_{\theta}$.

\item A lower bound of the form $\bar{y}_{n}^{\prime}/c$ reflects the
multiplicative model used to simulate controls; in contrast a lower bound of
the form $\bar{y}^{\prime}_n-b$ assumes an incompatible additive model.  \end{itemize}
By switching our attention from
$\theta^{\prime}$ to $\bar{\varepsilon}_{n}^{\prime}$, we decouple the performance of our procedure from the choice 
$\hat f_\theta$. That is, we achieve symmetry/invariance. The  duality between $\theta$ and $\bar{\varepsilon}_{n}$---and the
switching strategy it allows---lies at the heart of 
\textit{Fiducial inference} \citep{fisher1930,fisher1935},
which led to the development of confidence intervals in \citet{neyman1934}.
The general recipe may be summarized as follows. Identify a quantity $t\left(
\varepsilon^{\prime}\right)  $ with the following properties (in Example 7, $t\left(  \varepsilon^{\prime}\right)  =\bar{\varepsilon}%
_{n}^{\prime}$):

\begin{enumerate}
\item[I.] $t\left(  \varepsilon^{\prime}\right)  $ is dual to $\theta^{\prime}$,
i.e., inference for $t\left(  \varepsilon^{\prime}\right)  $ is equivalent to
inference for $\theta^{\prime}$. Rather than design procedures to
estimate $\theta^{\prime}$, we can design procedures to estimate
$t\left(  \varepsilon^{\prime}\right)  $.

\item[II.] The distribution of $t\left(  \varepsilon^{\prime}\right)  $ over
$\hat{\Omega}$ depends only on $\hat f_\varepsilon$ and 
\textit{not} on $\hat{f}_{\theta}$; such a quantity is said to be
\textit{pivotal}. It is usually 
simple to find ``estimators" of pivotal quantities whose
effectiveness is decoupled from our choice of $\hat f_\theta$. 
In our example, the value of $c$ which makes $\bar{\varepsilon}_{n}^{\prime}<c$ for $95\%$ 
of controls depends only on $\hat f_\varepsilon$.

\end{enumerate}
This approach gives us a systematic way of constructing symmetric/invariant 
procedures \citep[see][]{fraser1968,dawid1982,hannig2009}. 

\qquad\textbf{Pivots and the Relevance Robustness Tradeoff.} 
In principle, we can 
replace $\hat \Omega$ in Criterion II with $\hat \Omega_{\text{rele}}$; this 
lets us create procedures that are symmetric over \textit{subsets} of controls. 
However, we encountered a key phenomenon in Example 6: matching breaks symmetry/invariance. 
This implies that the more matching we do, the harder it will be to (non-trivially) satisfy Criteria I and II. Again, consider Example 7.

\begin{description}
\item[Case 1:] We perform no matching: $\hat{\Omega}_{\text{rele}}=\hat
{\Omega}$. The distribution of $\bar{\varepsilon}_{n}^{\prime}$ over
$\hat{\Omega}$ depends only on $\hat{f}_{\varepsilon}$ and not $\hat
{f}_{\theta}$; hence $\bar{\varepsilon}_{n}^{\prime}$ is pivotal.

\item[Case 2:] We use only completely matched controls, i.e., $y_{i}^{\prime
}=y_{i}$ for all $i$. The constraint $\bar{y}_{n}=\bar{y}_{n}^{\prime}%
=\theta^{\prime}\bar{\varepsilon}_{n}^{\prime}$ induces dependence between
$\theta^{\prime}$ and $\bar{\varepsilon}_{n}^{\prime}$ over the set of matched
controls. The distribution of $\bar{\varepsilon}_{n}^{\prime}$ among controls
with $\bar{y}_{n}^{\prime}=\bar{y}_{n}$ now depends on $\hat{f}_{\theta}$, i.e., 
$\bar{\varepsilon}_{n}^{\prime}$ is no longer pivotal \textit{with respect to
$\hat{\Omega}_{\text{rele}}$}. Our usage of the term ``pivotal" here is an extension of
the standard definition given above.
\end{description}

Whereas it is straightforward to find procedures with robust performance over
$\hat{\Omega}$, it is difficult to achieve similar robustness  over subsets of $\hat{\Omega}$. Most often, only
\textquotedblleft absurd\textquotedblright\ estimators can guarantee such robustness. 
For example, if
$\hat{\Omega}_{\text{rele}}$ is the set of complete matches in our current
example, only two (non-randomized) interval estimates exist whose performance
over $\hat{\Omega}_{\text{rele}}$ is insensitive to $\hat{f}_{\theta}$: the
interval, $\left(  0,\infty\right)  $, which is 100\% correct and 100\% unhelpful, or the empty set, 
which is clearly useless as well. 
\bigskip\qquad

\qquad\textbf{Creating Symmetry/Invariance  via Minimaxity. }Pivots are most useful when
constructing confidence intervals (or more generally, confidence sets) whose
performance is invariant to $\hat f_\theta$. We can try to apply the same logic to
symmetrize point estimators. This often works when a problem has inherent
group symmetries \citep[see][Chapter 3]{lehmann1998} but success is no longer guaranteed. 
This calls for a more general approach. As motivation, consider an asymmetric point
estimator. It is asymmetric because it is more accurate for some values of
$\theta$ than others. To make it more symmetric, we should \textquotedblleft
redistribute our point estimator's effort\textquotedblright\ from those values
of $\theta$ for which its risk is small to those where its risk is large, thus
making its risk as uniform in $\theta$ as possible$.$ Just as wealth
redistribution helps the poorest in society, this risk redistribution will
make the resulting estimator perform better in worst-case scenarios (but as a
tradeoff, worse in best-case\ scenarios).

\qquad The above reasoning hints that (approximately) symmetric  estimators are
those with good performance in worst case scenarios. Therefore, in our search
for symmetric estimators, we might focus attention on so-called
\textit{minimax} estimators, $\hat{\theta}_{\text{mini}}$, which minimize the
average error over $\hat{\Omega}$ under some ``least favorable" $f_{\theta}$
specification; see \citet{brown1994} and \citet{,strawderman2000}. However,
there are  caveats:

\begin{enumerate}
\item Minimaxity generates invariance  through redistribution which can lead to inefficiency.
 For example, suppose we want to estimate the probability, $\theta$, that a
coin lands heads using $n$ flips of the coin. Under squared error loss, the
fraction of heads out of $n$ flips (the sample mean) is not a minimax
estimator for $\theta$---it fares quite well when the true $\theta$ value is
near $0$ or $1$ but poorly when $\theta$ is near $\frac{1}{2}$. The actual
minimax estimator \textquotedblleft redistributes effort\textquotedblright%
\ towards $\frac{1}{2}$, giving itself a slight edge over the sample mean for
a small interval of $\theta$ values near $\frac{1}{2}$ but at a huge cost to
accuracy outside this interval
\citep[see][Chapter 5 Example 1.7]{lehmann1998}. In general,
\textit{engineered symmetry/invariance}, where none inherently exists, can lead to
serious efficiency loss across large regions of the parameter space. In
contrast, when a problem contains inherent group symmetries
\citep[][Chapter 3]{lehmann1998}, the redistribution strategy works quite well.

\item As with pivots, the more we match,
the harder it becomes to apply the minimax strategy.
 For example, minimax estimators over the set of complete matches
($D^{\prime}=D$) do not usually exist because the worst case loss is usually
infinity (or the maximum allowable loss) for \textit{any }estimator. Even with partial matching, the price of creating invariance may be too high \citep[see][]{brown1990}---this is simply
the relevance robustness tradeoff at play.
\end{enumerate}

\subsection{An Extreme Case of Asymmetry}

\label{sec:testing} 
\qquad Hypothesis testing problems often display a form of asymmetry so acute
that we cannot robustly assess test accuracy even when no matching is involved.
That is, we cannot reliably predict whether a hypothesis test
will give the right conclusion (at least not without external information).

\bigskip

\qquad\textbf{Example 8 (The Trouble with Testing).} Suppose noisy measurements,
$\left\{  y_{i}\right\}  _{i=1}^{n}$, are taken of the position of a star
along some axis, and  an additive model, $y_{i}=\theta+\varepsilon_{i}$, with
$\varepsilon_{i}$ treated as independent standard normal errors, is judged
appropriate. Astronomers wish to test the null hypothesis $H_{0}:\theta=0$
against the alternative hypothesis $H_{1}:\theta=\theta_{1}$ where $\theta
_{1}$ is an \textit{unknown}, \textit{non-zero} value. The canonical z-test
looks at the magnitude of the average position, $\left\vert \bar{y}%
_{n}\right\vert $. If the average position is far from $0$, say $\left\vert
\bar{y}_{n}\right\vert >1.96\sqrt{n}$, the test rejects $H_0$.
We examine the error rate of this test for various choices of $\hat{\Omega}$.

\begin{enumerate}
\item[Setting 1] For all control problems, set $\theta^{\prime}=0$ (i.e., we
specify $\hat{f}_{\theta}$ as a\textit{ \textquotedblleft}point
mass\textquotedblright\ at\textit{ }$0$). Our test commits an error if it
rejects the null. Under our normal model, $\bar{y}_{n}^{\prime}>1.96\sqrt{n},$
less than 5\% of the time when $\theta^{\prime}=0$. Thus our test has 5\%
error rate over $\hat{\Omega}$, its \textit{Type I error} (false positive rate).

\item[Setting 2] For all control problems, set $\theta^{\prime}=c$
(i.e. we specify $\hat{f}_{\theta}$ as a\textit{ \textquotedblleft}point
mass\textquotedblright\ at some\textit{ }non-zero $c$). Our test now
commits an error if it \textit{fails to reject} the null (a Type II error or
false negative). The error rate is highly dependent on our choice of the simulation
parameter $c$. As $\left\vert c \right\vert $ approaches $0$, the error rate
approaches 95\% (cf. Setting 1). For $\left\vert c\right\vert $
large, the error rate approaches 0\%. In practice, we can draw a \textit{power curve} (one
minus the error rate plotted as a function of $c$). 
\end{enumerate}
We see that the accuracy of the z-test  depends critically on our
specification of $\hat{\Omega}$; for arbitrary specifications of $\hat
{f}_{\theta}$, the error rate of our test can be as high as 95\% or as low as
0\%---an almost vacuous statement.

\qquad To understand the source of the asymmetry, note that we chose the
$1.96\sqrt{n}$ threshold so that our test would have 5\% error rate under
Setting 1. That is, hypothesis tests are
constructed so that they have guaranteed good performance for select choices
of $\hat{f}_{\theta}$. This \textquotedblleft favoritism\textquotedblright%
\ for select $\hat{f}_{\theta}$ does not usually exist in interval or point
estimation problems and is the reason why the real-life accuracy of a
hypothesis test is so hard to predict. The lower our error rate when
$\theta=0$, the higher it must be when $\theta$ is near but not equal to 0. To
predict the test's accuracy therefore requires prior information about the
magnitude of $\theta$; without this information, the Type I error and power
curve of a test---the two traditional measures of accuracy---do not allow us to
predict how well our test  will \textit{actually} perform.

\begin{remark}
Some of the asymmetry above can be alleviated by testing only for
\textquotedblleft interesting\textquotedblright\ alternatives. For example, we
may only care about \textit{practically different }deviations from $0$, say
$\left\vert \theta_{1}\right\vert \geq1$. With a sample size of $10$,
restriction of $\hat{f}_{\theta}$ to interesting alternatives bounds the error
rate of our test below 11.5\%; see \citet{johnson2010} and references therein
for more on this approach. 
\end{remark}

Except in special cases \citep[see][]{berger1994,berger2003}, inference for
the accuracy of a hypothesis test will not be robust to $\hat{f}_{\theta}$.
These special cases usually require us to treat false rejections on an equal
footing with failures to reject, alleviating the asymmetry in the problem.
There is hope, however. We are now in an era of massive parallel testing,
e.g., millions of genes are tested for association with disease risk.
Together, these parallel tests form an \textquotedblleft
empirical\textquotedblright\ set of control problems through which we can
assess a test's accuracy \citep[]{benjamini1995,efron2010}. As we will see in
Section \ref{sec:diagonal}, data structures like this open up new ways to
achieve relevance and robustness in testing.

\section{From the Bayes-Frequentist Dichotomy to the Relevance-Robustness Continuum}

\label{sec:reconcile}

\subsection{Let's Compromise: Partial Conditioning}

\label{sec:partial} 
A treatment behaves differently on different patients---this is the motivation
for \textit{personalized }medicine. Inspired by this idea, our central theme
has been conducting \textit{individualized} inference\textit{ }for a dataset.
For example, we can use  features of the data, via matching, to assess how well
the 5\% nominal error rate of a 95\% confidence interval applies to the
problem at hand. The success of individualization relies on our ability to
build controls which resemble our actual problem---this in turn relies on our
ability to meaningfully specify $\hat{f}_{\theta}$ and $\hat{f}_{\varepsilon}%
$. The greater the level of individualization we strive for, the more stress
we put on our model, the more fragile our inference. The degree of
individualization should therefore scale with the reliability of our model.
This logic underlies the hardest question in statistics: \textit{how
individualized }should (can) our inference be?

The difference between (subjective) Bayesian and Frequentist inference hinges on this
question. While they are often thought to be two different methodologies, in
fact they share the same logic, with the only difference being how they select
the relevant subset of control problems. What is usually dubbed as
Bayesian statistics, we  call full individualization (complete
matching); what is dubbed (unconditional) Frequentist statistics, we call no
individualization. We purposely eschew the traditional terminology to
highlight the fallacy of the dichotomy; Bayes and Frequentism are two ends of
the same spectrum---a spectrum defined in terms of relevance and robustness.
The nominal contrast between them---parameters are fixed for Frequentists but
random for Bayesians---is a red herring. With this in mind, we note that it
appears strange that the two most common answers to the question---how
individualized should we be---have also been the \textit{two most extreme
answers}. The appropriate tradeoff most likely lies in between, and will be
problem-dependent (the principle of individualization!). That is, a compromise
between Frequentist and Bayesian inference through partial matching may yield
a more satisfactory relevance robustness tradeoff.

\bigskip

\textbf{Example 9 (Partial Matching)}. Suppose we have background and outcome
information for $n$ individuals, $\{\vec{x}_{i},y_{i}\}_{i=1}^{n}$, and wish
to predict the outcome $y_{0}$ for a target individual with background
$\vec{x}_{0}$. Hence, $\theta=y_{0}$ and $D=\left\{  \vec{x}_{0},\{\vec{x}%
_{i},y_{i}\}_{i=1}^{n}\right\}  $. One possibility is to fit a regression
model $y_{i}=\vec{x}_{i}^{\top}\beta+\varepsilon_{i}$ via least squares, then
predict $y_{0}$ using $\hat{y}_{0}=\vec{x}_{0}^{\top}\hat{\beta}$, where
$\hat{\beta}$ is the regression estimate. To assess this procedure's accuracy,
we estimate its error, $\Delta=\hat{y}_{0}-y_{0}$, when applied to the target
individual. 

We simulate controls $\left(  D^{\prime},\theta^{\prime}\right)  $ using the
regression model $y_{i}^{\prime}=\vec{x}_{i}^{\top}\beta^{\prime}%
+\varepsilon_{i}^{\prime}$, with $\beta^{\prime}$ simulated according to a
distribution $\hat{f}_{\beta}$ and $\varepsilon_{i}^{\prime}$  as
independent standard normal errors. The prediction error for each
control problem decomposes into two terms
\begin{equation}
\Delta^{\prime}=\hat{y}_{0}^{\prime}-y_{0}^{\prime}=\vec{x}_{0}^{\top}%
(\hat{\beta}^{\prime}-\beta^{\prime})-\varepsilon_{0}^{\prime
}\text{.}\label{e:bayes}%
\end{equation}
Regardless of which aspects of the data we match on and regardless of our
choice for $\hat{f}_{\beta}$, the distribution of the second term $\varepsilon_{0}^{\prime
}$ is always
standard normal. Thus, in deciding the appropriate degree of matching, it
suffices to study the first term. Consider now three choices for the set of relevant
controls, $\hat{\Omega
}_{\text{rele}}$, occupying three positions on the relevance robustness spectrum.

\begin{itemize}
\item[I.] \textit{Complete Matching (Bayes): }$D^{\prime}=D$. Due to matching,
$\hat{\beta}^{\prime}=\hat{\beta}$ for all controls in $\hat{\Omega
}_{\text{rele}}$. The first term in (\ref{e:bayes}) becomes $\vec{x}^{\top}(\hat{\beta}%
-\beta^{\prime})$; its distribution over the set of complete matches (known as
the posterior distribution) is sensitive to $\hat{f}_{\beta}$. If we trust $\hat f_\beta$,
however, it is the ``optimal" inference (Section \ref{sec:analogy}).

\item[II.] \textit{No Matching (Frequentist)}: $\hat{\Omega}_{\text{rele}%
}=\hat{\Omega}$. Without matching, the first term in (\ref{e:bayes}) becomes a pivotal
quantity, i.e., its distribution over $\hat{\Omega}$ is independent of $\hat
{f}_{\beta}$.  
\end{itemize}

Finally, we strike a compromise by conducting our \textit{error assessment} on
\textquotedblleft partially\textquotedblright\ matched controls. For algebraic
simplicity, we assume below that individual 1 and the target individual have
the same background: $\vec{x}_{1}=\vec{x}_{0}$. Disclaimer: we chose this scheme because the resulting inference, (\ref{e:key}), cleanly
illustrates how partial matching constitutes a
\textquotedblleft compromise\textquotedblright\ inference. This was a
pedagogical \textit{not }a practical choice.

\begin{itemize}
\item[III.] \textit{Partial Matching: }$D^{\prime}$ \textit{matches }$D$
\textit{in all aspects except }$y_{1}^{\prime}$ \textit{may not equal }$y_{1}%
$. Over this subset of partial matches, it can be shown that the first
term in (\ref{e:bayes}) becomes
(see Remark 5 below)\begin{equation}
\vec{x}_{0}^{\top}(\hat{\beta}^{\prime}-\beta^{\prime})=(1-h_{0})\cdot
B^{\prime}+h_{0}\cdot F^{\prime}\label{e:key}%
\end{equation}
where $0\leq h_{0}\leq1$ acts as a weighting factor.  The distribution of $B^{\prime}$ over $\hat \Omega_{\text{rele}}$ turns out to be very close to the posterior distribution formed under Strategy I. \textit{If}
$\hat{f}_{\beta}$ is correctly specified, it gives a near optimal characterization of our
uncertainty regarding $\Delta$; as with the Bayes inference above, it is relevant but not robust. The
distribution of $F^{\prime}$ over $\hat{\Omega}_{\text{rele}}$ is independent
of $\hat{f}_{\beta}$ (cf. Strategy II); it gives a robust but less relevant inference.  Partial matching creates
a compromise inference by using a weighted average of $B^{\prime}$ and $F^{\prime}$!
\end{itemize}

\bigskip

\begin{remark}
The weight $h_{0}$ equals $\vec{x}_{0}^{\top}(X_{n}^{\top}X_{n})^{-1}\vec
{x}_{0}$ where $X_{n}$ is the usual design matrix formed from $\{\vec{x}%
_{i}\}_{i=1}^{n}$; it is often called the leverage. The term  $B^{\prime}$ equals
$\vec{x}_{0}^{\top}(\hat{\beta}-\beta^{\prime}),$ which coincides with our
error term in the complete matching analysis (note that unlike for fully
matched controls, $\hat{\beta}^{\prime}\neq\hat{\beta}$ for partially matched
controls). The term $F^{\prime}$ equals $\vec{x}_{0}^{\top}(\hat{\beta}^{\prime}%
-\beta^{\prime})+(\Delta_{1}-\Delta_{1}^{\prime})$; its distribution over the
set of partial matches is free of $\hat{f}_{\beta}$. Here $\Delta_{1}=\hat
{y}_{1}-y_{1}$ is the \textquotedblleft prediction error\textquotedblright%
\ for the first individual (using the least squares estimator) and $\Delta
_{1}^{\prime}$ is the analogous quantity for control problems. 
\end{remark}

\bigskip

Partial matching gives us a flexible way to vary the degree of relevance versus robustness
in  our inference. Complications do exist \citep[see][]{berger1988}---most
salient, which features should we match/condition on and which should we
ignore? There is no unique best answer. However, this does not imply that we
should abandon matching in general or always match on everything---the doctor
in her treatment of Mr. Payne has certainly not adopted such radical
positions! Even though the \textit{theory} of partial matching is not even
partially complete, partial matching can nonetheless serve as a useful tool in
\textit{practice} as long as we  understand its aim: to assess whether the
uncertainty associated with \textit{our }problem should be greater (data are
\textquotedblleft unlucky\textquotedblright) or less (data are
\textquotedblleft lucky\textquotedblright) than the uncertainty of the\textit{
typical} problem in $\hat{\Omega}$. The goal is to identify features of the
data that  capture aspects of data-luck that are sufficiently robust to
$\hat{f}_{\theta}$. As long as one keeps this principle in mind, one does not
have to be bound by the dogma of Frequentism \textit{or }Bayes. One can be
Frequentist, Bayesian, and many other things, including being a Fiducialist
(see Section \ref{sec:diagonal}).

Whereas there is no generic method for deciding the best relevance robustness tradeoff, we can offer a whole host of examples.
Figure \ref{fig:venn} depicts a Venn diagram---references on the left give
examples of inferences that veer too far towards robustness, and those on the
right too far towards relevance. The goal is to stand in the middle, in
the \textquotedblleft football\textquotedblright\ (a concept developed by Carl
Morris and Joe Blitzstein for their inference course at
Harvard). Hopefully, the reader can map their own problem in terms of these \textquotedblleft
reference\textquotedblright\ points.

\begin{figure}[ptbh]
\begin{center}
\includegraphics[width=0.75\textwidth]{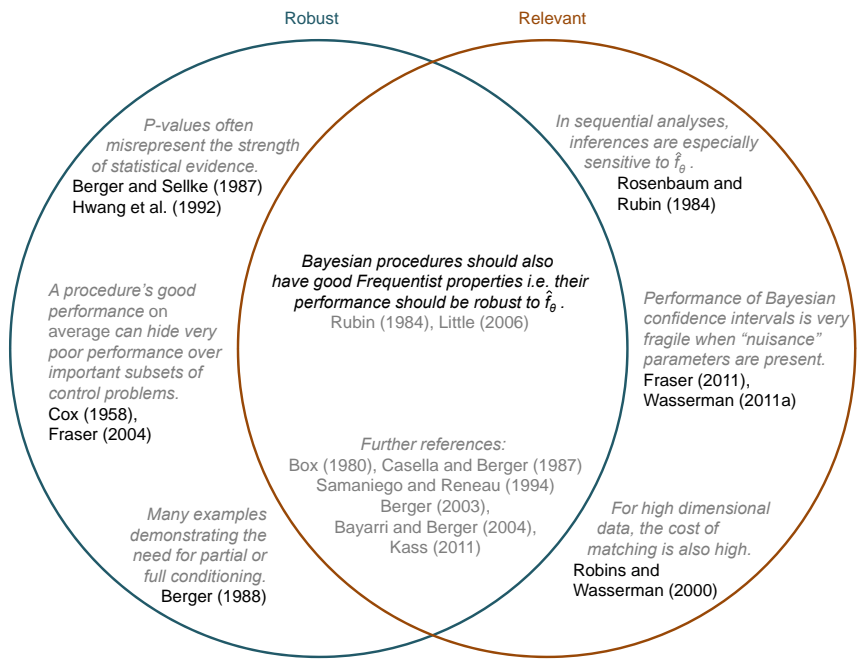}
\end{center}
\caption{What constitutes a ``good" procedure for \textit{our} problem? There are two basic criterion: (a) when we evaluate its performance in a highly individualized manner (i.e., using only controls which closely match our problem), the procedure should have good performance for some reasonable choice of $\hat f_\theta$ and (b) when we evaluate its
performance in a less individualized manner, the procedure should have good performance across a broad range of $\hat f_\theta$ specifications. Such procedures can be found in the intersection of the Venn
diagram. References in the non-intersecting regions discuss the pitfalls of
dogmatically skewing in favor of relevance or robustness.}%
\label{fig:venn}
\end{figure}

\subsection{The Feasibility Diagonal}

\label{sec:diagonal}

\begin{figure}[ptbh]
\begin{center}
\includegraphics[width=1\textwidth]{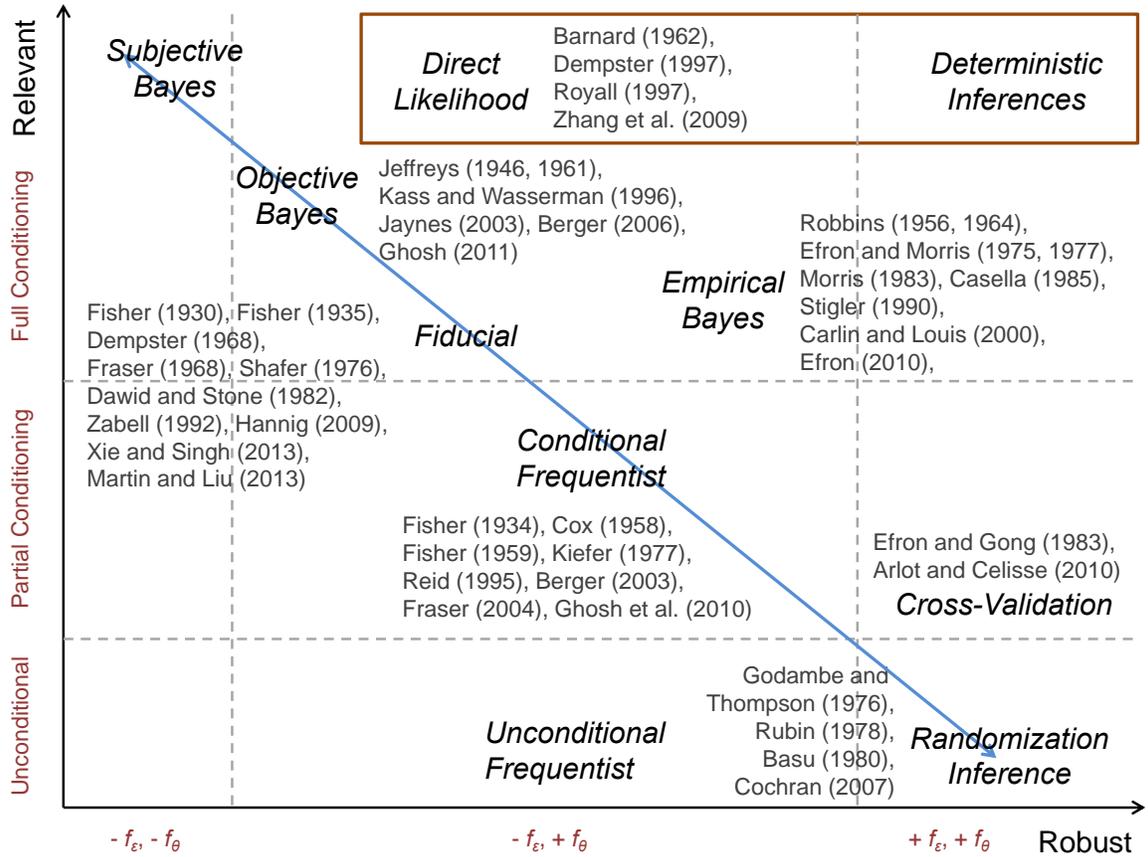}
\end{center}
\caption{Methods of inference are mapped according to their (robustness,
relevance) coordinates. $+$ denotes robustness to modeling of $f_{\theta}$ or
$f_{\varepsilon}$ and $-$ denotes a lack of such robustness. The diagonal line is the feasibility diagonal. To move above
the diagonal requires additional assumptions or sacrifices. The boxed region in the top-right corner
depicts methods which give non-probabilistic inferences.}
\label{fig:inftable2}
\end{figure}

Our central point is that there is no \textquotedblleft
right\textquotedblright\ inference, \textit{only tradeoffs}. Figure
\ref{fig:inftable2} maps several common statistical methodologies based on the
relevance robustness tradeoff they make. The blue line marks the
\textit{feasibility diagonal}---it represents the total budget we can spend on
relevance or robustness. At one end of the diagonal lies subjective Bayesian
inference, the optimal allocation if we completely trust the information
encoded in $\hat{f}_{\theta}$ and $\hat{f}_{\varepsilon}$. At the other end
lies randomization inference (e.g., the Fisher Randomization test; see
\citealt{basu1980}) whose validity is free of both $\hat{f}_{\theta}$ and
$\hat{f}_{\varepsilon}$. In between lies the objective Bayesian, the
Fiducialist, and the conditional Frequentist (A caveat: we have grouped under
\textquotedblleft Fiducial" several different methodologies whose
reasoning process share deep connections with Fisher's original development).
Historically, these three approaches arose out of attempts to create an
\textquotedblleft objective\textquotedblright\ inference. Many of these
efforts ran into trouble because defining \textquotedblleft
objectivity\textquotedblright\ is highly non-trivial (if not impossible)! By
situating these three developments along the feasibility diagonal, we may
reinterpret the historical quest for \textquotedblleft
objectivity\textquotedblright\ as a quest for an optimal relevance robustness
tradeoff. Our hope is that the latter perspective offers a more tractable
formulation for the problem of inference. Along these lines, one should view
objective Bayes, Fiducial and conditional Frequentism as promising initial
starting points (having been beta-tested on canonical examples) in our search
for a satisfactory balance between relevance and robustness. But ideally, we
will adapt these initial proposals to the special features of the problem at
hand. That is, we will follow David Cox's advice that there are no routine
statistical questions but only questionable statistical routines, and discover 
problem specific sweet spots along the feasibility diagonal.

\textbf{Moving off the feasibility diagonal}. It is trivial to move below the
feasibility diagonal (a suboptimal position) by failing to match on important
ancillary features of the data (see Example 2). But can we move above the
diagonal, to achieve increases in both relevance and robustness? There is of
course no free lunch---to move above the diagonal, either a sacrifice must be
made (e.g., in terms of interpretability or scope of applicability) or else an
assumption must be added. We consider four important examples: 

\begin{itemize}
\item \textit{Deterministic Inferences}: Statistical inference is
overwhelmingly probabilistic because in most situations it is impossible to
make non-trivial deterministic statements. However, as \citet{wasserman2011a}
points out, there do exist important applications
\citep[for example, see][]{cesa2006} where deterministic control of prediction
error can be achieved. 
\item \textit{Direct Likelihood Inference }%
\citep[]{barnard1962,royall1997}: Likelihood is a way of measuring
the degree to which the data supports various hypotheses or parameter values.
Importantly, the likelihood depends only on our specification of $\hat
{f}_{\varepsilon}$ and not of $\hat{f}_{\theta}$---thus the allure of drawing
inferences directly from the likelihood. \citet{mealli2009} give an example
where this strategy may be appropriate. The sticking point is a
lack of interpretability, i.e.,   ``likelihood units" do not have an operational/real life meaning.
Thus, while direct likelihood methods (and deterministic inferences) score
high on both relevance and robustness, they score low on a third dimension---\textit{utility}---
which could not be visualized in our 2-D figure. Remember: no free lunch.
 \item \textit{Cross Validation }\citep{efron1983,arlot2010}: Previously 
we used the data to build a simulation model for control problems, $\left(
D^{\prime},\theta^{\prime}\right)  $. Certain applications allow for a more direct
approach to creating controls by partitioning the original data as $D=\left(
D^{\prime},\theta^{\prime}\right)  $. The $D^{\prime}$ created in this way is
known as the \textquotedblleft training set\textquotedblright, $\theta
^{\prime}$ as the \textquotedblleft test set\textquotedblright. Consider
Example 9, where we wish to assess the least squares prediction error for a
target individual with background $\vec{x}_{0}$. Suppose as before, $\vec
{x}_{0}=\vec{x}_{1}$. Hence the first individual is a good proxy for the
target individual.  This suggests a
\textquotedblleft leave-one-out\textquotedblright\ strategy: fit a regression
model using only individuals $2$ to $n$ (the training set) and compute the
error from using the fitted model to predict the first individual's response
(the test set). This error is an estimate of our prediction error for the
target individual; it can be a rather noisy estimate because we use a
control population of size 1 (the first individual). To reduce our estimate's variability, we might apply the leave-one-out strategy to
\textit{each} of the $n$ individuals in our dataset, i.e., we predict each individual's response using the remaining $n-1$ individuals and then average the resulting errors. This effectively gives us a control population of size $n$. In comparison to Example 9 where we simulated controls
using a regression model, the leave-one-out strategy does not require an
explicit model---our inference will be more robust.
The downside is that we base our accuracy assessment on observed
prediction errors for individuals who do not necessarily match the target's background
$\vec{x}_0$---our inference will be less relevant.

\item \textit{Empirical Bayes }\citep{robbins1956,efron1977,casella1985}:
Certain data structures allow for a reliable specification of $\hat{f}%
_{\theta}$. Consider Example 6 where the doctor wants to predict your disease
status, $\theta$, based on the result of a test, $y$. In its original form,
the problem favors no matching because the doctor may not know
 the proportion of sick individuals in the population.  
But now suppose the doctor has an
additional source of information: test results $\left\{  y_{i}\right\}
_{i=1}^{n}$ from $n$ of his other patients.  This allows us to accurately estimate the prevalence,  at least when $n$ is large:   if  the prevalence is low/high,
there should be proportionally more negative/positive results among the $n$ patients.
Empirical Bayes refers to methods which exploit such forms of indirect information
\citep[]{efron2010} to reliably deliver highly individualized inferences---here, the test results for the \textit{other} $n$ patients
provide indirect information about \textit{your }disease status.
\end{itemize}

Whereas the relevance robustness tradeoff is inescapable, these examples
demonstrate how the particularities of a problem might allow us to recast the
exact terms of the tradeoff. It is through such tinkering that innovations in
inference take place. Figure \ref{fig:inftable2} depicts the world of inference---relevance and
robustness are its longitude and latitude. We were only able to hit the
tourist traps on this trip, but we hope the reader, armed with an
understanding of relevance and robustness, will
return to navigate Figure \ref{fig:inftable2} at their own leisure.
 
\section*{Epilogue: A Rebirth for Individualized Inference?}

Standard statistical procedures guarantee good
performance \textit{on average} over some set of controls; they are targeted
at the ``typical" control problem. But every problem is \textit{atypical}
in a different way. Just as the doctor must judge whether a clinically successful
treatment will continue to be successful given Mr.\ Payne's unique background, 
so the data analyst must gauge how well a procedure's nominal
guarantee, e.g., 95\% confidence or 5\% Type I error, applies to \textit{this
dataset}. Individualized inference aims to deliver uncertainty assessments
that are relevant for the problem at hand.

\textbf{Is Individualization Even Possible? }To develop a personalized
treatment, we need to conduct personalized experiments, which requires
personalized guinea pigs. The heart of statistical inference lies in \textit{designing} 
relevant control problems. Many mistake the probabilistic statements statisticians make as
``natural properties" and statisticians as their ``discoverers." Nothing could be further from the truth.
Probability does not simply exist, it is \textit{given
existence }through judgements about which control problems are and are not relevant for the actual problem. 
This point also reveals the risk of individualization---if our judgement is poor, our inferences will be worse off. 
The quality of our model limits the degree of individualization we can reasonably achieve---to go beyond
this limit is to inject misinformation. Divisions in the statistical community
center around this question: what is the \textit{appropriate degree}?
Frequentists and Bayesians stand at the extremes of no individualization and
complete individualization, but the degree of
individualization should vary with the problem. The extent to which a doctor
personalizes treatment depends on her understanding of the disease and
treatment; rather than any fixed relevance robustness tradeoff, the analyst must adapt her
inference to the problem environment. Those arguing against individualization
may deem such adaptation \textquotedblleft ad hoc\textquotedblright\ and
\textquotedblleft subjective\textquotedblright---this \textit{mistakes
insensitivity for objectivity}.

\textbf{Why Individualization Matters More Today. }A more apt term for Big
Data is perhaps complex data. With increasingly complex structures, the ways a
dataset can be \textit{atypical} multiply. The statistical models of the past
often produced relatively \textit{homogeneous} control
problems. The costs of individualization often outweighed its benefits. To
accommodate the complex features of modern data, statistical models today
inevitably produce ever more \textit{heterogeneous} sets of controls
(\citealt{efron2007} provides a striking example in the context of microarray
data). The persuasiveness of guarantees on \textit{average performance}
deteriorates with increased heterogeneity (see Example 5). Hence the need to
deliver \textit{individualized assessments of uncertainty} is more pressing than ever.

\section*{Acknowledgements}

Deeply insightful comments from Jessica Hwang, Paulo Orenstein, Robin Gong, Andrew Gelman, Iavor Bojinov, Guillaume Basse, and William Meeker improved this paper greatly. The reader has them to thank
for any clarities and us to thank for all the inclarities. Keli is indebted to Joe
Blitzstein and Carl Morris for their impassioned teaching; they made
statistical inference into a beautiful way to reason rather than a series of
theorems and formulas. An earlier version of this paper served as the basis
for a presentation at the First International Workshop on Bayesian, Fiducial
and Frequentist Inference (BFF) held in Shanghai; we appreciate the
feedback and encouragements from all the BFFs (Best Friends Forever) in
attendance. We also thank the editor Stephen Stigler for his comments and his extraordinary
patience as we completed this work. Finally, we thank the NSF and JTF for partial
financial support.

\bibliographystyle{plain}
\bibliography{inference_final}
\nocite{*}
\end{document}